%% file: sample-acmtog-SIGGRAPH-submission.tex
\documentclass[acmtog, nonacm]{acmart}
\acmSubmissionID{118}

\usepackage{booktabs} 

\usepackage{color} 

\usepackage[ruled]{algorithm2e} 

\SetAlFnt{\small}
\SetAlCapFnt{\small}
\SetAlCapNameFnt{\small}
\SetAlCapHSkip{0pt}

\setcopyright{rightsretained}
\acmJournal{TOG}
\acmYear{2024} \acmVolume{43} \acmNumber{4} \acmArticle{} \acmMonth{7}\acmDOI{10.1145/3658120}

\begin{document}
\title{SketchDream: Sketch-based Text-to-3D Generation and Editing}

\author{Feng-Lin Liu} 
\affiliation{
\institution{Institute of Computing Technology, CAS and University of Chinese Academy of Sciences}
\country{China}
}

\author{Hongbo Fu}
\affiliation{
\institution{SCM, City University of Hong Kong, and EMIA, HKUST}
\country{China}
}
\email{fuplus@gmail.com}

\author{Yu-Kun Lai}
\affiliation{
\institution{School of Computer Science and Informatics, Cardiff University}
\country{UK}
}
\email{LaiY4@cardiff.ac.uk}

\author{Lin Gao}
\authornote{Corresponding author: Lin Gao (gaolin@ict.ac.cn).}
\affiliation{%
\institution{Institute of Computing Technology, CAS and University of Chinese Academy of Sciences }
\country{China}
}
\email{gaolin@ict.ac.cn}

\authorsaddresses{Authors' addresses: Lin Gao and Feng-Lin Liu are with the Beijing Key Laboratory of Mobile Computing and Pervasive Device, Institute of Computing Technology, Chinese Academy of Sciences, and University of Chinese Academy of Sciences. Hongbo Fu was with the SCM, City University of Hong Kong, and is now with EMIA, HKUST. Yu-Kun Lai is with the School of Computer Science and Informatics at Cardiff University. 
Authors' e-mails: liufenglin21s@ict.ac.cn, fuplus@gmail.com, LaiY4@cardiff.ac.uk, gaolin@ict.ac.cn.
\\This is the author's version of the work. It is posted here for your personal use. Not for redistribution. }

\newcommand{\sysName}{SketchDream}

\begin{teaserfigure}
  \centering
  \includegraphics[width=1\linewidth]{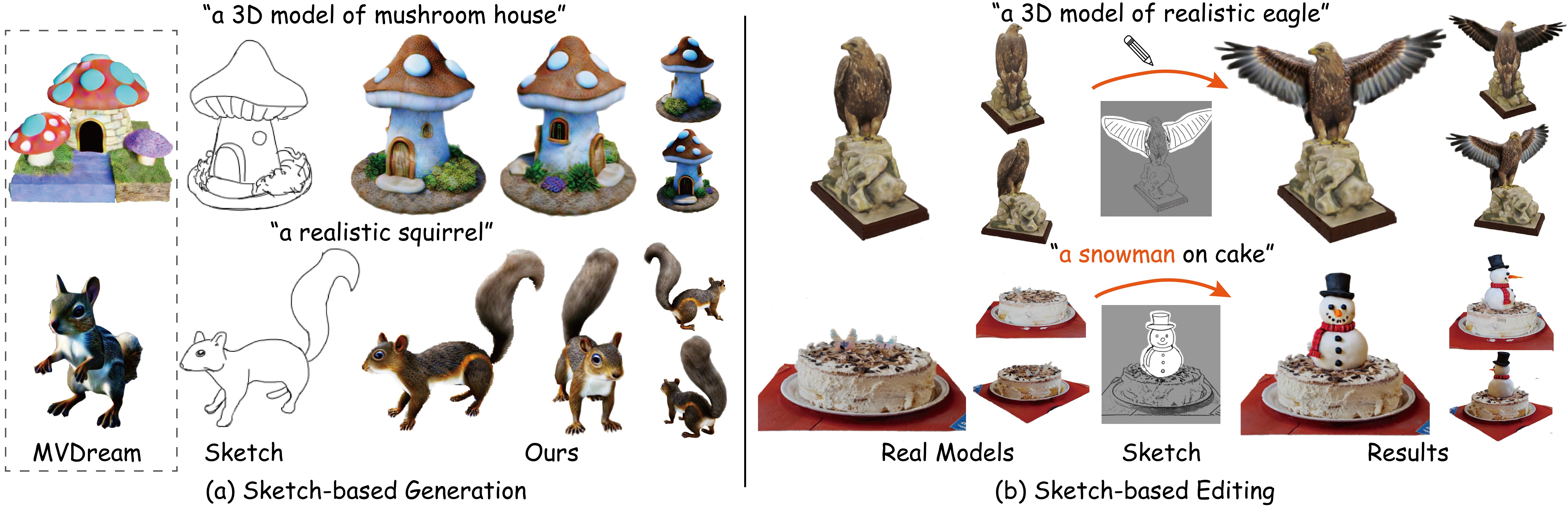}
  \caption{Our {\sysName} system supports both generation and editing of high-quality 3D contents from 2D sketches.
  As shown in (a), given hand-drawn sketches and text prompts (on top of each example), our method generates high-quality rendering results of 3D contents from scratch. 
  Existing text-to-3D generation approaches like MVDream \cite{mvdream} generate photo-realistic results but cannot control component layouts and details, such as the door and window in the top example and the pose in the bottom example. 
  In (b), we show sketch-based editing results of NeRFs reconstructed from real models.
  The newly generated components naturally interact with the original objects, with the unedited regions well preserved. 
  }
  \label{fig:teaser}
\end{teaserfigure}

\begin{abstract}
Existing text-based 3D generation methods generate attractive results but lack detailed geometry control. 
Sketches, known for their conciseness and expressiveness, have contributed to intuitive 3D modeling but are confined to producing texture-less mesh models within predefined categories. 
Integrating sketch and text simultaneously for 3D generation promises enhanced control over geometry and appearance but faces challenges from 2D-to-3D translation ambiguity and multi-modal condition integration.
Moreover, further editing of 3D models in arbitrary views will give users more freedom to customize their models. However, it is difficult to achieve high generation quality, preserve unedited regions, and manage proper interactions between shape components. 
To solve the above issues, we propose a text-driven 3D content generation and editing method, {\sysName}, which supports NeRF generation from given hand-drawn sketches and achieves free-view sketch-based local editing. 
To tackle the 2D-to-3D ambiguity challenge, we introduce a sketch-based multi-view image generation diffusion model, which leverages depth guidance to establish spatial correspondence. 
A 3D ControlNet with a 3D attention module is utilized to control multi-view images and ensure their 3D consistency. 
To support local editing, we further propose a coarse-to-fine editing approach:
the coarse phase analyzes component interactions and provides 3D masks to label edited regions, while the fine stage generates realistic results with refined details by local enhancement. 
Extensive experiments validate that our method generates higher-quality results compared with a combination of 2D ControlNet and image-to-3D generation techniques and achieves detailed control compared with existing diffusion-based 3D editing approaches. 

\end{abstract}



\maketitle

\input{sections/1_intro}
\input{sections/2_relate_work}

\input{sections/3_preliminary}

\input{sections/4_method}

\input{sections/5_exp}

\input{sections/6_conclusion}

\bibliographystyle{ACM-Reference-Format}
\bibliography{sample-bibliography}

\end{document}

%% file: sections/1_intro.tex
\section{Introduction}

Creating high-quality 3D content is a popular topic with wide applicability in VR/AR, the movie industry, architecture, robotics simulation, etc. 
However, traditional 3D content production depends on sophisticated software and laborious procedures, making it challenging for amateur users to design their 3D models. 
To solve this problem, sketching, a user-friendly and expressive interaction tool, has been utilized in 2D \cite{Sketch2photo, pix2pix, controlnet} and 3D content generation \cite{xiang2020sketch_mesh, gao2022_sketchsampler, zhang2021sketch2model, zheng_diffusion_SDF, SketchFaceNeRF}. 
However, due to the restriction of network capacity and the ambiguity of 2D sketches, existing works can only generate limited single or multiple categories of objects. 
Additionally, since sketches only contain geometry information, how to effectively control the appearance has been under-explored in existing sketch-based 3D generation works. 

Compared with sketches, text prompts can describe object category and appearance more easily.
Thanks to the development of diffusion models, text-to-image generation \cite{stable_diffusion, Midjournal} has become successful in recent years. 
Based on these pre-trained 2D models, DreamFusion \cite{Dreamfusion} designs Score Distillation Sampling (SDS) to optimize Neural Radiance Fields (NeRFs) and presents a text-to-3D generation framework. 
Follow-up works \cite{magic3d, Fantasia3D, richdreamer, ProlificDreamer, mvdream} further modify the supervision losses and optimization process to generate more realistic and higher-quality 3D models. 
Despite impressive results of text-based 3D generation, text prompts cannot precisely depict objects' shape, texture patterns, and layouts, and are thus less suitable for fine-grained control. 
Moreover, it is difficult for the above works to edit the local details of generated or real 3D models because of their global control of textual description. 
Although some works~\cite{vox-e, Dreameditor, progressive3d} further achieve text-based local editing by labeling editing regions, it is still hard for them to easily control the shape and position of edited components. 

To improve the controllability of 3D generation, we introduce sketch into the text-to-3D generation framework: sketch guides the shape and pattern, while text prompt controls the material and appearance. 
This goal is challenging to achieve for the following reasons. First, since a sketch only contains single-view information, the absence of multi-view supervision makes it difficult to generate complete 3D models.
Moreover, directly adding single-view sketch constraints in other viewpoints degrades the generation quality and has lower faithfulness. 
For example, a straightforward approach \cite{IPDreamer} is to first generate 2D images by ControlNet \cite{controlnet} with a sketch-based condition and then utilize image-to-3D approaches \cite{magic123, zero123, Dreamcraft3d} to generate 3D contents. 
However, as shown in Fig. \ref{fig:generation_compare}, this approach tends to generate distorted geometry in novel views different from input sketches and fuzzy texture details in the back view due to the information absence.

Further editing of the generated contents in other views provides users with more detailed control of results. Additionally, editing existing real models instead of generation from scratch is also a common situation. 
However, sketch-based 3D content editing is challenging because besides achieving high-quality content creation, the edited content should reasonably interact with the original content, with unedited regions well preserved.
To solve the above issue, SKED~\cite{Sked} requires the input of multi-view sketches, which are treated as binary masks to label edited regions. 
Although it achieves effective text- and sketch-based editing at the object level, maintaining unedited regions well and supporting part-level editing are still challenging.

To address the above issues, we propose {\sysName}, a method for sketch-based text-to-3D generation and editing of photo-realistic contents.
For sketch-based generation, since sketches are sparse and lack 3D information, it is ambiguous and difficult to directly generate 3D contents from them. 
To solve this problem, we use the depth information, which can bridge the 2D inputs and 3D models. 
Specifically, given an input 2D sketch and a text prompt, we utilize a 2D diffusion model \cite{controlnet} to generate a corresponding depth map, which is then utilized to warp the input sketch. 
To propagate the sketch into 3D space and avoid the Janus problem (i.e., 3D models with multiple frontal faces), we build a 3D ControlNet based on MVDream \cite{mvdream} to generate images in four camera views uniformly distributed in azimuth.
To build the correspondence between the sketch and multi-view images, we modify the MVDream backbone to generate an additional image in the view of the input sketch.
To ensure the consistency between different views and support sketch control, we design a 3D-attention control module, which takes the input 2D sketch and warped sketch in the nearby view with depth guidance as inputs to effectively control the 3D diffusion generation. 
Similar to MVDream \cite{mvdream}, we use SDS-based optimization to generate high-quality 3D contents.

Our framework naturally supports sketch-based 3D content editing. 
Given the generated or real 3D models, users can input an edited sketch with a 2D editing mask to modify local components. 
To generate high-quality editing results, we design a two-stage coarse-to-fine editing framework. 
In the coarse stage, the 2D mask is lifted into 3D space to construct a coarse 3D columnar mask. Then, the coarse stage generates an editing result that generally conforms to the edited sketch and the input text, but might have unoptimized quality, inadequate faithfulness, and mistaken interaction with the original object. 
We optimize the editing result quality in the fine stage.
In particular, we extract a mesh model from the coarse-stage editing result and then label the edited region by the coarse 3D columnar mask.
This new 3D mask precisely represents the component interaction and helps preserve the unedited regions.
Moreover, we propose a local rendering strategy that calculates the sketch-based SDS supervision locally to enhance the sketch faithfulness and generation quality further.

Extensive experiments validate that our method generates higher-quality results than possible sketch-based text-to-3D baselines and existing sketch-based 3D editing approaches.
Our main contributions can be summarized as follows.
\begin{itemize}
\item We propose the first sketch-based text-to-3D generation and editing method, which generates high-quality 3D objects under generalized categories and supports detailed editing of reconstructed or generated NeRFs. 
\item We propose a sketch-based multi-view image generation diffusion model, which utilizes a depth-guided warping strategy to create spatial correspondence, and a 3D-attention control module to ensure 3D consistency. 
\item To support local modification further, we develop a coarse-to-fine editing framework. The coarse stage generates initial results to label edited regions better, while the fine stage generates high-quality editing results with a local rendering strategy.
\end{itemize}

%% file: sections/2_relate_work.tex
\section{Related work}

\paragraph{Sketch-based 3D Generation}
Sketch-based 3D generation has been extensively researched. 
Early works utilize retrieval strategies \cite{2003_search_sketch_generation, 2003visual_retrieval} or carefully designed mapping approaches \cite{igarashi2006teddy, zeleznik2006sketch} to determine 3D shapes by sketches.  
With the development of deep learning, recent works treat this problem as sketch-conditioned 3D reconstruction and predict volumetric grids, point clouds, mesh models, or even CAD commands \cite{li2020sketch2cad, li2022free2cad}. 
Although volumetric prediction methods \cite{delanoy2018_3d_volumn, 2018_unsupervised_volumn} generate 3D models faithful to sketches, the grid resolution restricts the model quality. 
To enhance the details, another category of works \cite{lun2017_3d_pc, wang2022_3d_pc, gao2022_sketchsampler} map sketches into point clouds. 
To further construct 3D shape meshes, other works \cite{li2018robust_surface, xiang2020sketch_mesh, zhong2020towards} predict the depth and normal maps, which are utilized to deform {pre-defined templates} or directly construct 3D shapes. 
Data augmentation \cite{zhong2020deep_data_aug} and sketch preprocessing \cite{zhang2021sketch2model} are also utilized to improve the robustness of generation from hand-drawn sketches. 
Despite the successful generation results, the above approaches are hard to achieve detailed local control for out-of-domain examples. 
Zheng et al.~\shortcite{zheng_diffusion_SDF} propose a two stage diffusion with local attention mechanism to generate SDF models. 

Different from the above works, our method utilizes NeRF as a 3D representation and thus synthesizes 3D shapes with realistic textures instead of geometry models only. 
Besides, our method is not restricted to single or multiple object categories and aims for a more general framework for sketch-based 3D model generation. 

\paragraph{Text- and Image-based 3D Generation}
With the development of diffusion models, text-based 3D content generation has become popular in recent years. One category of methods (e.g., \cite{Rodin}) utilizes diffusion models to generate 3D representation{s} like tri-plane features directly, but their performance and generalization are limited by 3D training datasets. 
Another category of methods employs pre-trained 2D diffusion models \cite{stable_diffusion} to optimize 3D representations by the SDS loss \cite{Dreamfusion} or SJC (Score Jacobian Chaining) loss \cite{SJC}. 
The subsequent works try to improve the performance based on the above framework. 
Some works divide the generation process into two stages: geometry optimization and texture optimization, and utilize diverse 3D representations like DMTet \cite{magic3d}, BRDF \cite{Fantasia3D}, and Gaussian Splatting \cite{Dreamgaussian}. 
Other works modify the supervision during generation by designing, for example, latent space optimization \cite{latent_nerf}, VSD {(Variational Score Distillation)} loss \cite{ProlificDreamer}, interval score matching \cite{luciddreamer}, and normal-depth supervision~\cite{richdreamer}. 
MVDream \cite{mvdream} further proposes a four-view generation model and solves the Janus problem. 
Please refer to insightful surveys \cite{survey_3D_Sun0024, survey_3D_XiaX24, survey_3D_Liu24} for comprehensive understanding of the 3D generation approaches.
Although the above works generate high-quality results, users cannot precisely control the geometry shapes and texture details.
Compared with the above works, our method adds easily drawn sketches as an additional condition and achieves more detailed control during the text-based generation. 

To improve the controllability of 3D generation, many works utilize single-view images to replace or combine with text prompts as input.
Make-It-3D \cite{Make-It-3D} and RealFusion \cite{Realfusion} add image constraints to 3D generation, but {still fail to} generate {optimal} results in the back view. 
Follow-up works \cite{zero123, zero123++, SyncDreamer} modify Stable Diffusion {\cite{stable_diffusion}} to infer other views of a single-view image.
Their pre-trained models are utilized in Magic123 \cite{magic123} to achieve better geometry results in novel views. 
DreamCraft3D \cite{Dreamcraft3d} and HyperDreamer \cite{Hyperdreamer} respectively introduce DreamBooth \cite{Dreambooth} and image super-resolution to enhance the texture quality. 
Concurrently, ImageDream \cite{Imagedream} adds image condition into a multi-view generation model to improve the generation robustness. 
However, in real applications, obtaining a desirable 2D image with detailed control is nontrivial. 
Additionally, it is difficult for the above works to support local modification because of the global dependence on the input images. 
Apart from real images, Control3D \cite{control3d} utilizes sketches as conditions for diffusion-based 3D generation but tends to generate fuzzy details due to the dependence of 2D ControlNet. 
A concurrent work of our method, MVControl \cite{mvcontrol} extracts edge maps from images as an additional condition, but do not deal with hand-drawn sketches and local editing. 
In contrast, our method liberates the reliance on images, enabling high-quality 3D generation from scratch, and additionally facilitates sketch-based local editing.

\begin{figure*}
    \centering
    \includegraphics[width=1.0\linewidth]{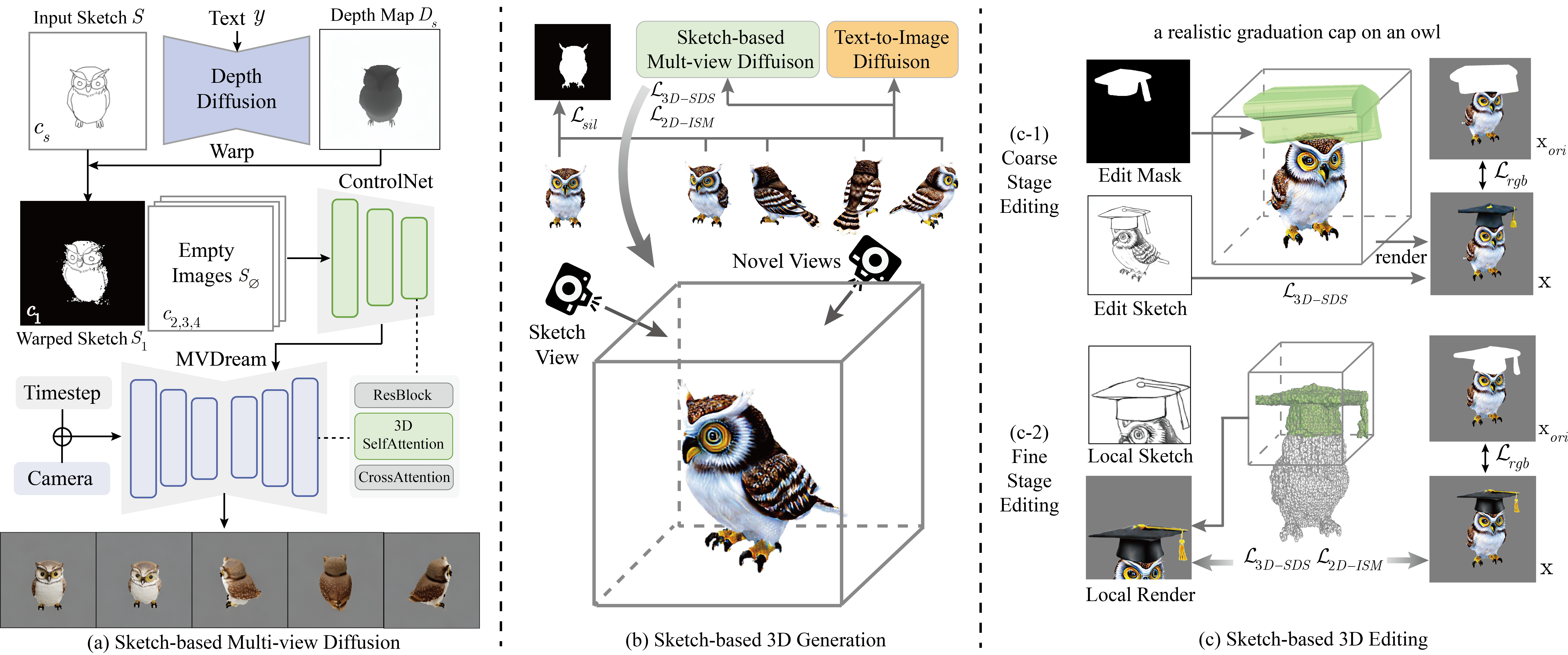}
    \caption{
    The overview of our {\sysName} for sketch-based generation and editing. 
    Given an input sketch $S$ and a text prompt $y$, we design a sketch-based multi-view diffusion model (a), which takes {$S$}, depth-warped sketch $S_1$, and white images $S_{\varnothing}$ as condition{s} {and} generates multi-view images in the sketch view $c_s$ and novel view{s} $c_{mv}$. 
    In order to generate realistic 3D contents (b), we render images under five views corresponding to those in the multi-view diffusion model and then optimize a NeRF by 3D score distillation and 2D interval score matching. 
    For sketch-based 3D editing (c), we design a two-stage editing framework. 
    In the coarse stage, we build a coarse 3D mask and generates a coarse editing result, which is used to get precise 3D masks for high-quality local editing in the fine stage. 
    }
    \label{fig:pipeline}
\end{figure*}

\paragraph{3D Content Editing}
Compared with 3D generation from scratch, 3D content editing requires additional considerations on the relationship between the edited and original contents and the preservation of unedited regions.
Previous methods only support geometry editing \cite{xu2022deforming, nerf_editing, VolTeMorph, SketchFaceNeRF}, require laborious interaction operations \cite{Neumesh, yang2021learning, Edit_NeRF}, or only support global manipulation \cite{editable_nerf, stylenerf, SNeRF}. 
To achieve more detailed and semantic-driven editing, many approaches \cite{clip-nerf, TextDeformer, nerf-art} utilize the pre-trained CLIP model \cite{CLIP} to compute the text guidance loss, which limits their performance because of CLIP's training for text-image alignment instead of generation. 
To further improve the editing performance, the subsequent works utilize Stable Diffusion for 3D editing. 
These approaches utilize a user-defined 3D geometry box \cite{progressive3d, focaldreamer} or an inferred 3D mask by multi-modal attention \cite{Dreameditor, vox-e} to label local editing regions, supporting editing manipulations like adding or replacing objects while preserving unedited regions in scenes. However, it is still difficult to control the shape or position of the edited content precisely by using text prompts. 
Most similar to our work, SKED \cite{Sked} also supports sketch-based control of 3D editing results. 
However, SKED requires multi-view sketches as input and cannot control texture details because of the transformation of sketches into texture-less masks.
Compared with existing works, our method supports single-view sketch-based editing and achieves more complicated editing operations like controllable geometry modification.

%% file: sections/3_preliminary.tex
\section{Preliminary}\label{sec:Preliminary}

\paragraph{MVDream.}
To avoid the Janus problem and generate correct geometry, we build our framework on MVDream \cite{mvdream}. 
With text inputs, MVDream simultaneously generates images under four views, which have uniformly distributed view angles, with the same elevation. 
To control view angles, the absolute camera extrinsic matrix is encoded and added to the time-step embedding in a UNet.
To ensure cross-view consistency, MVDream utilizes a 3D attention module that shares the queries $Q$, keys $K$, and values $V$ in all views.
The diffusion model is fine-tuned from 2D Stable Diffusion on the Objaverse \cite{objaverse} dataset. 

\paragraph{Score Distillation Sampling (SDS).}
First proposed in DreamFusion \cite{Dreamfusion}, SDS has been widely used in text-to-3D generation based on 2D diffusion models. 
To mitigate the color saturation problem, we utilize an SDS version with $x_{0}$-reconstruction loss:
\begin{equation}
    \mathcal{L}_{SDS}(\theta,\rm{x}=g(\theta, c)) = {\mathbb{E}}_{t,c,\epsilon}[\left \| \rm{x}-\hat{\rm{x}}_{0} \right \|_{2}^{2} ],
    \label{eq:eq1}
\end{equation}
where $g(\theta, c)$ is a rendered image of 3D representation $\theta$ with camera condition $c$. 
$\hat{\rm{x}}_{0}$ is the estimated $\rm{x}_{0}$ {(images without noise)} 
based on the {UNet's} output $\epsilon_{\phi}({\rm{x}}_{t}; y,c,t)$, where $y$ is a text condition, $x_{t}$ is the diffusion forward process results at time step $t$ with Gaussian noise $\epsilon$. 
In each optimization step, $t$ and $\epsilon$ are randomly sampled, constraining the rendered images $\rm{x}$ {to satisfy} the distribution of pre-trained diffusion models $\phi$. 
More details can be found in \cite{mvdream}.

%% file: sections/4_method.tex
\section{Methodology}

In this section, we present our framework for sketch-based text-to-3D content generation and editing, as illustrated in Fig. \ref{fig:pipeline}. 
Our approach addresses the challenge of synthesizing realistic 3D contents from sparse 2D sketches and textual inputs by completing missing appearance details and extending single-view information into the 3D space. 
To accomplish this, we introduce a sketch-based {multi-view diffusion} model in Sec. \ref{sec:method_multi-view}. This model predicts depth maps for sketch warping to establish spatial correspondence and generates realistic multi-view images. 
We detail our sketch-based 3D content generation to achieve a seamless 3D representation in Sec. \ref{sec:method_Generation}. 
The sketch-based multi-view diffusion model collaborates with SDS to apply 3D constraints and sketch control. Simultaneously, a pre-trained 2D text-to-image diffusion model enhances appearance details. 
Sec. \ref{sec:method_Editing} delves into our sketch-based editing framework, showcasing its ability to facilitate effective local editing while preserving the original features in unedited regions.

\subsection{Sketch-based Multi-View Diffusion Model}\label{sec:method_multi-view}
To achieve sketch-based 3D generation, the missing texture and material should be added, and the geometry information contained in 2D sketches should be propagated into 3D space. 
To achieve these goals, we design a sketch-based multi-view image generation model, which is built on the pre-trained MVDream~\cite{mvdream} because of its powerful 3D understanding ability.
To achieve sketch control, we modify the MVDream backbone to generate four novel-view images with an additional image in the sketch view.
Formally, given an input sketch $S$ with a corresponding camera condition $c_s$ and a text condition $y$, our multi-view diffusion model generates a sketch-view image ${\rm{x}}_{s}$ and novel-view images 
$\{{\rm{x}}_{1},\rm{x}_{1},\rm{x}_{3},\rm{x}_{4}\}$
corresponding to {novel-view} camera conditions $\{ c_{1}, c_{2}, c_{3}, c_{4}\}$.
We design a depth-guided warping strategy to explicitly build the spatial correspondence and utilize a 3D attention module to ensure 3D consistency. 

\begin{figure*}[]
    \centering
    \includegraphics[width=0.97\linewidth]{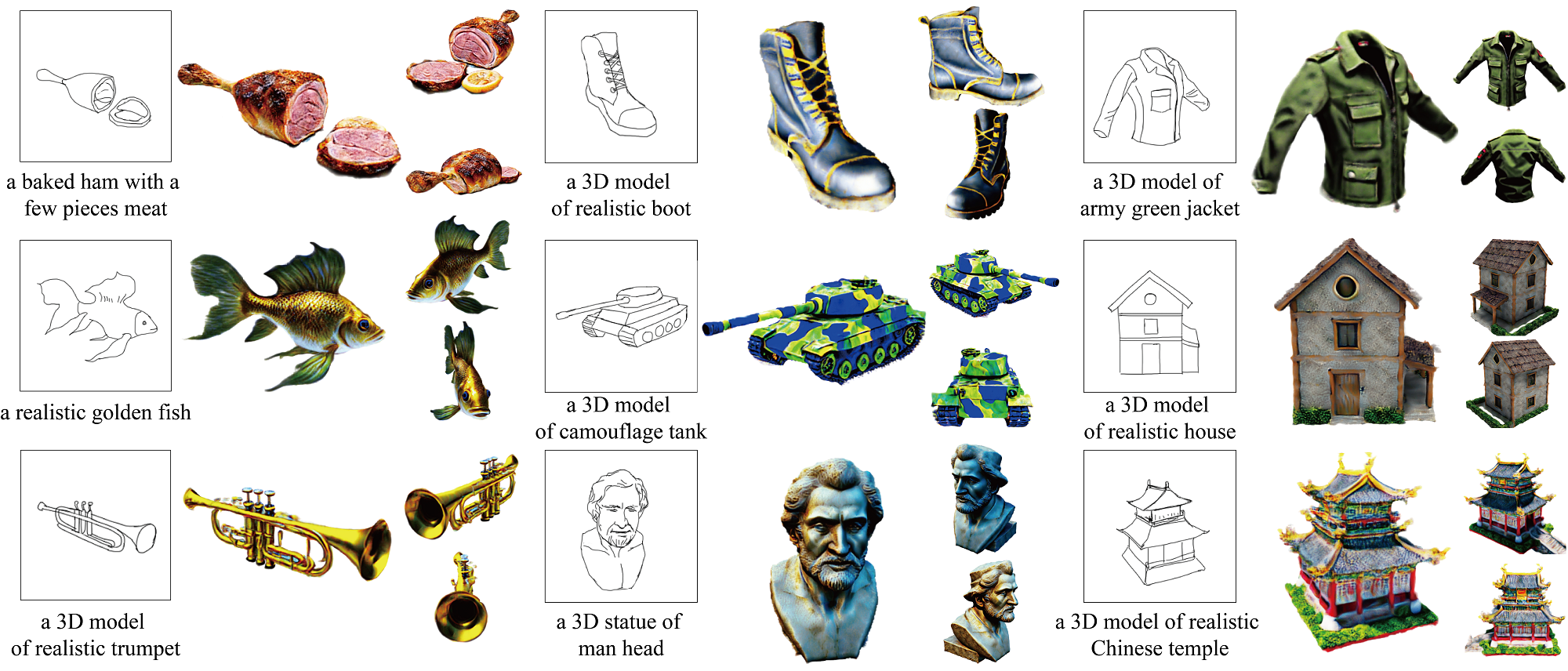}
    \caption{ 
    Sketch-based generation results. Given hand-drawn sketches, our method generates high-quality 3D results, which are faithful to the input sketches and texts. 
    Our method can generate models under diverse categories, including clothes, food, animals, humanoid objects, etc. 
    It can been seen that the shape and pattern details can be controlled by sketches. 
    }
    \label{fig:geneation_results}
\end{figure*}

\paragraph{Depth-guided Warping.} 
Adding an additional control like sketch into pre-trained 2D diffusion models has become successful in previous works~\cite{controlnet, T2I-Adapter}.  
However, different from 2D image translation with precise pixel alignment, sketch-based novel view image synthesis requires a complicated understanding of 3D geometry to generate correct results.
We observe that depth maps can serve as an intermediate geometry representation to solve the sketch's ambiguity and improve the faithfulness of sketches to generated models. 

We build a depth map generation diffusion model, which takes the sketch $S$ and text $y$ as input and generates the corresponding depth map $D_{s}$. 
Since the success of 2D ControlNet comes from the spatial alignment, we build this correspondence in 3D space by depth-guided pose warping~\cite{fehn2004depth, posewarping}.
Specifically, given the source depth $D_{s}$ and camera parameters $c_s$, we reproject the pixels of {the sketch image} into 3D space. 
Instead of warping the input sketch $S$ to all four novel views, we only warp it to the nearest view $c_{1}$ because the quality of the warped sketch in a far viewpoint can degrade significantly.
The warped target sketch $S_{1}$ is generated by interpolating the source pixel values. 
Formally, we denote the depth-guidance warping process as:
\begin{equation}
    S_{1} = Warp(S, D_s, c_s, c_1).
\end{equation}

\paragraph{3D Attention Control Module.}
We design a specific 3D attention control module to apply ControlNet \cite{controlnet} to the pre-trained MVDream. 
In the sketch view $c_s$, the input of the control branch is the sketch $S$ itself. 
For novel view images, the input in the nearest view $c_1$ is the warped sketch $S_1$, while the inputs of the other views $\{c_{2},c_{3},c_{4}\}$ are empty images $S_{\varnothing}$.
To ensure 3D consistency, the 3D ControlNet utilizes the 3D attention module \cite{mvdream}. In this module, as discussed in Sec. \ref{sec:Preliminary}, the Q, K, and V features in the self-attention layers are shared across multiple views, thus generating consistent multi-view images.
It should be noticed that although the inputs of the ControlNet in far views are empty, the 3D attention shares the sketch condition information to achieve effective control. 

\paragraph{Training.}
We trained the depth generation model and the sketch-based multi-view generation model separately. 
The former was finetuned based on the 2D ControlNet.
During the training, we fixed the VAE and ControlNet branch and trained the text-to-image UNet.
We treated the depth map as a color image and set the resolution as 256 $\times$ 256.
To train the networks, we utilize a subset of Objaverse \cite{objaverse} containing 150k 3D objects with 24 rendered images for each object, wherein 8 images with random viewpoints serve as input views and the other 16 images serve as target images. 
For the images of input views, we extract paired sketches by the method in \cite{Line_drawing}.

To train the multi-view ControlNet, we utilized the zero convolution to initialize the newly added layers as in the 2D ControlNet.
Given the dataset of paired multi-view images and sketches, the training samples $\{{\rm{x}}, s, y, c\}$ contain images {$\rm{x}={\{\rm{x}}_s, \rm{x}_{1},\rm{x}_{2},\rm{x}_{3},\rm{x}_{4}\}$}, sketch conditions $s=\{S, S_1, S_{\varnothing}\}$, {text conditions $y$}, and camera conditions {$c=\{c_s, c_{1}, c_{2}, c_{3}, c_{4}\}$}. 
We define the controllable multi-view diffusion loss as:
\begin{equation}
    \mathcal{L}_{MV-Ctrl}(\phi) = {\mathbb{E}}_{{\rm{x}},t,y,s,c,\epsilon}[\left \| \epsilon-\epsilon_{\phi}({\rm{x}}_t;t,y,c,s)) \right \|_{2}^{2} ],
\end{equation}
where $t$ is {a} time step {and} ${\rm{x}}_{t}$ is the noisy image generated from $\rm{x}$. 

\subsection{Sketch-based 3D Generation}\label{sec:method_Generation}
Since the generated 4-view images in Sec.~\ref{sec:method_multi-view} are view-sparse and not strictly 3D consistent, it is hard to directly optimize NeRF. Additionally, randomly sampling the 4 orthogonal views $N$ times generates {4$\times N$} images, which, however, might have different geometry, color, and texture, thus lacking 3D coherence.
Therefore, we utilize SDS optimization to generate 3D contents.
Specifically, we render five view images $\rm{x}$: the sketch view image to control the geometry and four randomly sampled view images to optimize the 3D NeRF representation. 
We calculate the 3D SDS $\mathcal{L}_{SDS}^{3D}$ as defined in Equation \eqref{eq:eq1} while {utilizing} the sketch-based diffusion network $\epsilon_{\phi}({\rm{x}}_t;t,y,c,s))$. 
Similar to MVDream \cite{mvdream}, we use the classifier free guidance (CFG) rescale trick to mitigate the color over-saturation. 

Apart from the 3D SDS constraint, we also utilize the 2D pre-trained text-to-image model (Stable Diffusion~\cite{stable_diffusion}).
We use the Interval Score Matching (ISM) \cite{luciddreamer} loss, which is more robust and generates more realistic results in our framework than the original SDS. 
To further improve the sketch faithfulness, we apply a 2D silhouette loss in the sketch view:
\begin{equation}
    \mathcal{L}_{sil} = \left \| M_s-C_s^\alpha \right \|_{2}^{2}, 
\end{equation}
where $C_s^\alpha$ is the rendered {object alpha} mask by 3D NeRF, and $M_{s}$ is the sketched region obtained by background removal.
The overall objective of our 3D generation process is:
\begin{equation}
    \mathcal{L}_{total}(\theta) = \lambda_{1} \mathcal{L}_{SDS}^{3D} + \lambda_{2} \mathcal{L}_{ISM}^{2D} + \lambda_{3} \mathcal{L}_{sil} + \lambda_{4} \mathcal{L}_{orient}, 
\end{equation}
where $\mathcal{L}_{orient}$ is the regular orientation loss proposed by \cite{Dreamfusion}.
We also turn on the point lighting \cite{Dreamfusion} and soft shading \cite{magic3d} as in MVDream to smooth the geometry. 

\subsection{Sketch-based 3D Editing}\label{sec:method_Editing}
Our framework supports sketch-based local editing of the generated or reconstructed NeRFs of 3D models.
To achieve detailed control, users are provided with sketches~\cite{Line_drawing} synthesized from rendered images. Subsequently, they can modify these sketches and draw an additional mask to label edited regions. 
Since it is challenging to directly infer the interaction between object components and generate high-quality results, we design a two-stage editing strategy. 
In the coarse stage, we get an initial editing result, which is used to predict a detailed 3D mask representing the component interaction. 
In the fine stage, the framework generates realistic, high-quality editing results while precisely preserving the unedited regions.

\begin{figure}[h]
    \centering
    \includegraphics[width=0.97\linewidth]{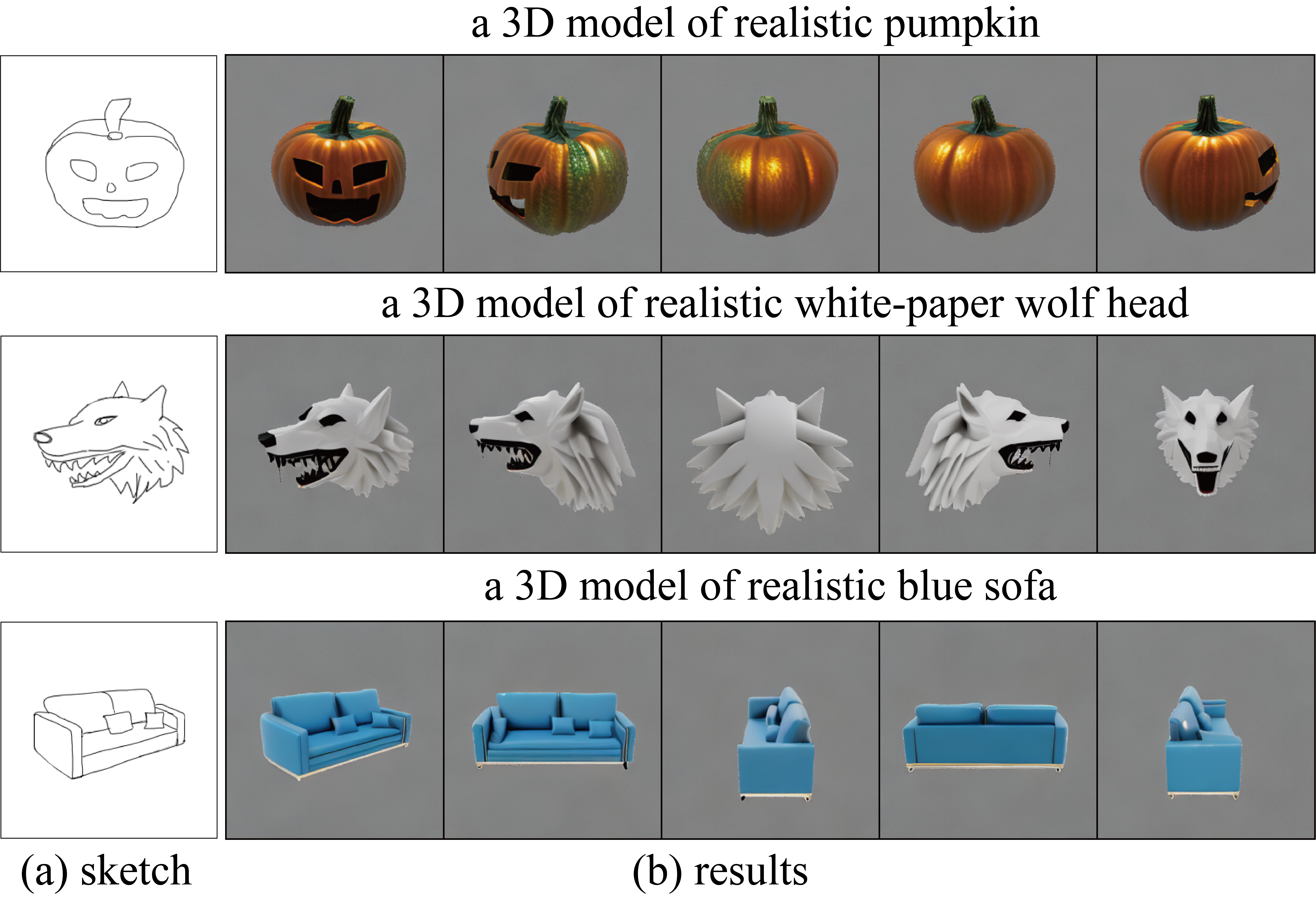}
    \caption{
    Sketch-based multi-view image generation results. 
    Given hand-drawn sketches (a) and texts shown above the images, our method generates realistic multi-view images (b), which are faithful to the input sketches and text prompts. 
    }
    \label{fig:geneation_results_sample}
\end{figure}

\subsubsection{Coarse-Stage Editing}
To enable effective 3D local editing, our approach involves the transformation of a 2D mask into 3D space to label editing regions in novel views. 
In the coarse stage, we utilize a cylinder mesh model for this purpose.
Beginning with a hand-drawn 2D mask, users define the minimum and maximum depth values, which are utilized to construct the top and bottom surfaces of the cylinder. 
Similar to sketch-based generation in Sec. \ref{sec:method_Generation}, we render five view images $\rm{x}$ to optimize the NeRF. 
To maintain the unedited regions, {we render} the 3D mask model in the camera conditions $c$ to generate $M_{2D}$. 
We define an image loss to preserve the color features:
\begin{equation}
    \mathcal{L}_{rgb} = \left \| {\rm{x}} \odot \overline{M}_{2D} - {\rm{x}}_{ori} \odot \overline{M}_{2D} \right \|_{2}^{2}, 
\end{equation}
where ${\rm{x}}_{ori}$ are {the multi-view rendered images before editing}.
Since we only need to get coarse editing results in this stage, the 2D loss is not used. The overall objective is:
\begin{equation}
    \mathcal{L}_{total}^{coarse}(\theta) = \alpha_{1} \mathcal{L}_{SDS}^{3D} + \alpha_{2} \mathcal{L}_{rgb} + \alpha_{3} \mathcal{L}_{sil} + \alpha_{4} \mathcal{L}_{orient}.
\end{equation}

\begin{figure}
    \centering
    \includegraphics[width=0.97\linewidth]{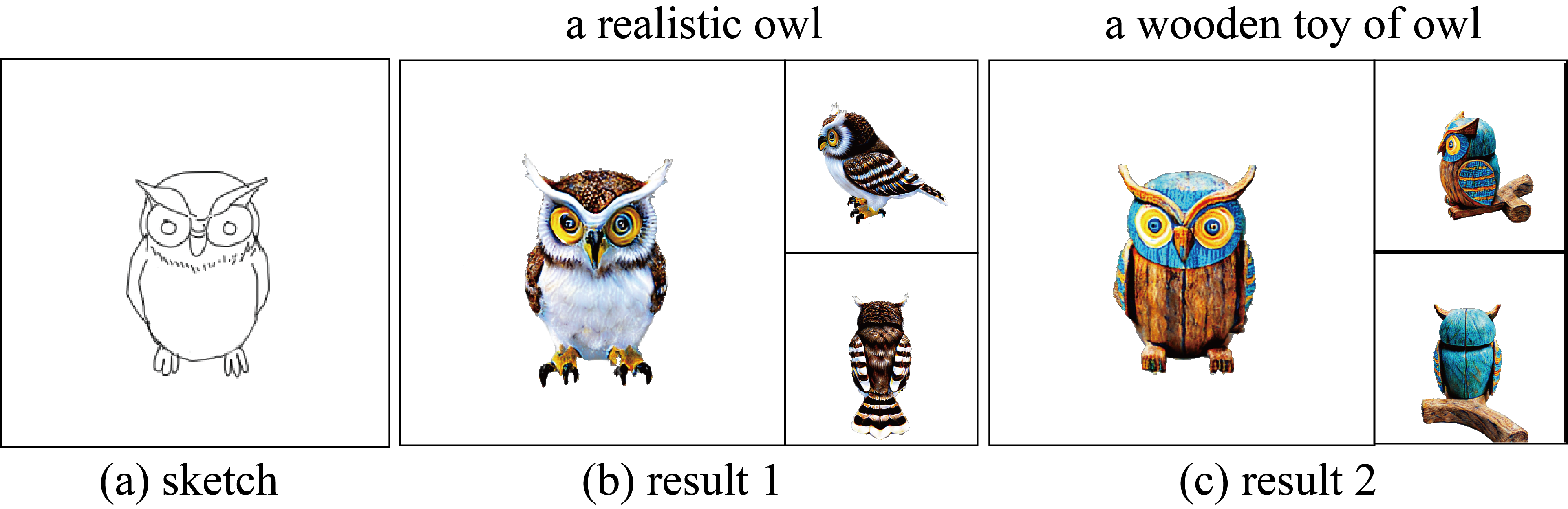}
    \caption{
    Sketch-based generation results with different text prompts. 
    Our method generates diverse and realistic results, whose geometry is controlled by the input sketch while appearance {being controlled by the text prompts}. 
    }
    \label{fig:different_text}
\end{figure}

\begin{figure*}[h]
    \centering
    \includegraphics[width=0.97\linewidth]{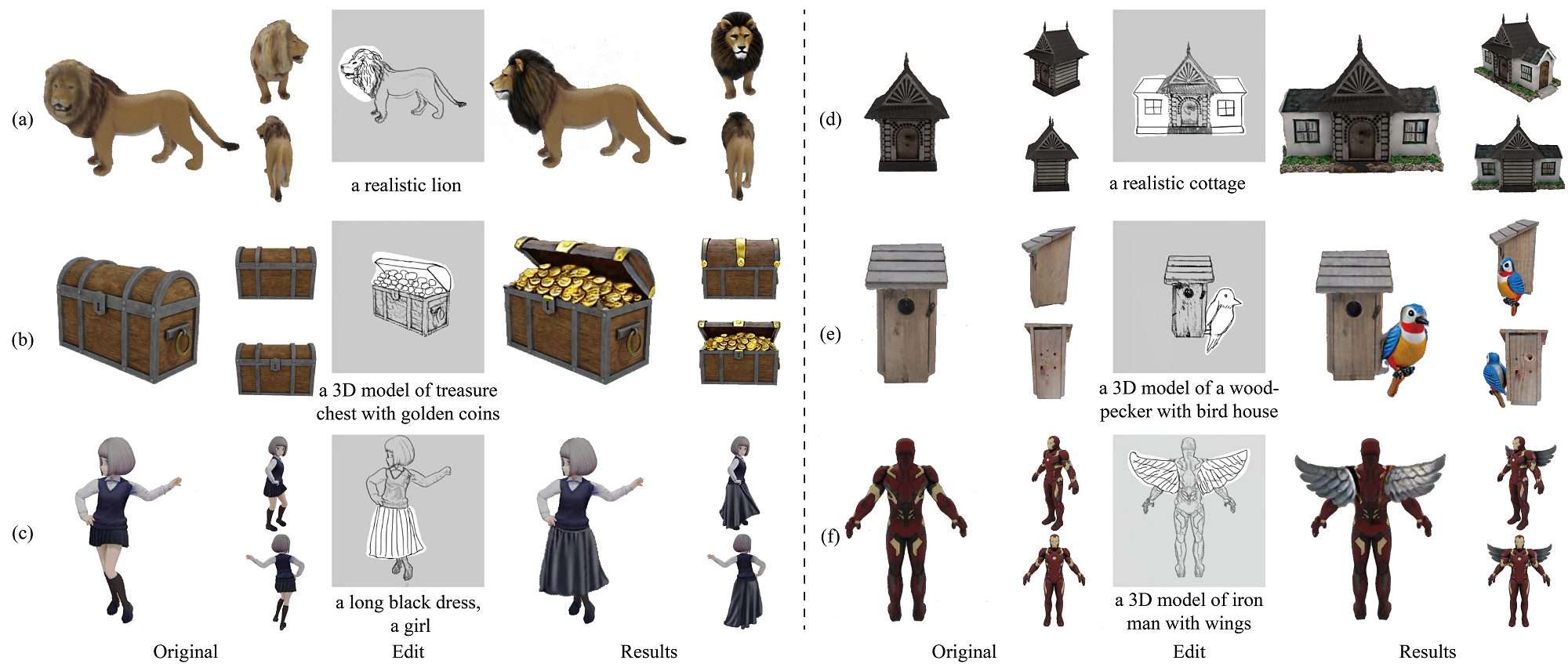}
    \caption{
    Sketch-based editing results. 
    Given real 3D models, users can select arbitrary views to render images {and} edit the local regions by inputting texts, modifying sketches (extracted from original render images), and drawing masks. 
    Our method {supports the change of} local components of real models, such as changing the lion's head {orientation} (a) opening the treasure chest (b), and changing clothes (c).
    Our method also supports adding new high-quality components with natural interactions with the original components.
    }
    \label{fig:results_editing}
\end{figure*}

\subsubsection{Fine-Stage Editing}
Utilizing the coarse 3D mask generates an initial editing result, but some undesirable regions are also mistakenly included or undesirably changed, as shown in Fig. \ref{fig:ablation_editing}. 
To address this issue, we get a more precise 3D mask to improve the editing performance based on the results of the coarse stage. 
Since NeRF lacks segmentation information and is computationally expensive, we translate the coarse NeRF results into a mesh model.
This 3D model has the newly generated components and generally maintains the original geometry in the unedited regions. 
Then, we label local regions of the 3D mesh to represent the edited regions by setting the vertices within the coarse mesh as the editing vertices, followed by manual refinement. It is convenient to paint on the mesh model by modifying vertex colors.
In the case of editing/removing existing components, we extract the mesh model from the original NeRF and apply the same labeling strategy, unifying the results with the newly generated one.
During the optimization, we render the newly labeled 3D mask regions to generate a precise 2D mask $M_{2D}$, which is used to calculate $\mathcal{L}_{rgb}$ to preserve the unedited regions. 

Globally optimizing the whole edited object can correctly determine the interaction between object components. However, we find that it tends to generate fuzzy results and have low faithfulness to the input sketches. 
To further improve the editing performance, we propose a local enhancement strategy that separately adds diffusion constraints in the local editing regions. 
Specifically, we construct the bounding sphere of the refined 3D mask. 
The sphere center defines the camera viewpoint while its radius defines the camera position. 
We utilize the local camera parameters to render the local editing regions, focusing the network attention into the interested components. 
During editing optimization, we randomly render the global images or local regions and optimize the following objective: 
\begin{equation}
    \mathcal{L}_{total}^{fine}(\theta) = \beta_{1} \mathcal{L}_{SDS}^{3D} + \beta_{2} \mathcal{L}_{ISM}^{2D} +  \beta_{3} \mathcal{L}_{rgb} + \beta_{4} \mathcal{L}_{sil} + \beta_{5} \mathcal{L}_{orient}.
\end{equation}
We utilize the 2D diffusion loss in {the} fine stage to improve the details. 
$L_{rgb}$ is calculated similarly to {that in} the coarse stage, with $M_{2D}$ rendered by the precise 3D mask.

%% file: sections/5_exp.tex
\section{Evaluation}

In this section, we conduct a series of qualitative and quantitative experiments to demonstrate the superiority of our framework to the alternative solutions. 
In Sec. \ref{sec:results}, we present the sketch-based generation and editing results. 
In Sec. \ref{sec:Comparison}, we show a comparison with state-of-the-art methods. 
In Sec. \ref{sec:ablation}, we conduct {an} ablation study to validate the effectiveness of the key components of our framework.
We also conduct a user study in Sec. \ref{sec:user_study} to further support the better performance and interaction of our approach. 

\begin{figure}
    \centering
    \includegraphics[width=0.97\linewidth]{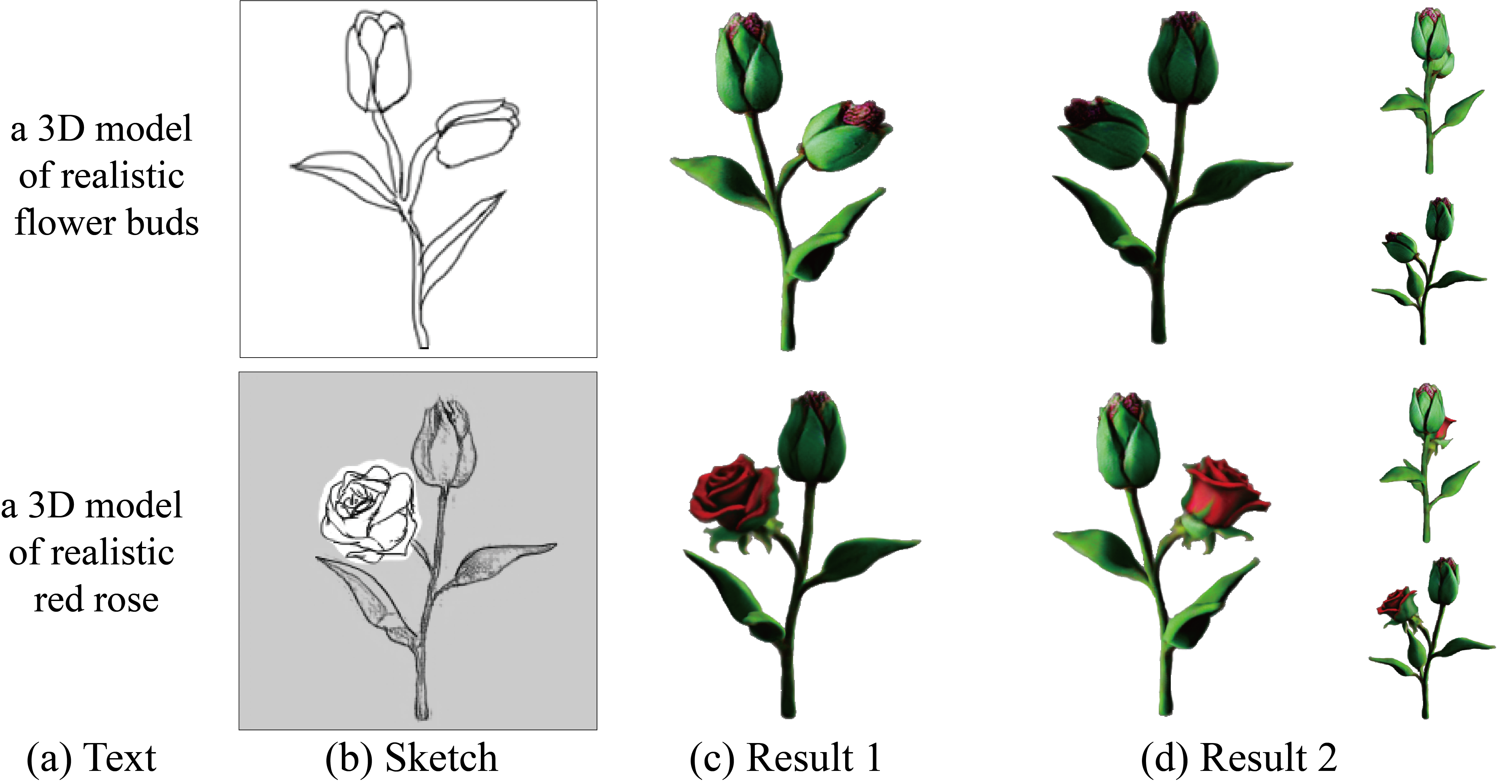}
    \caption{
    Sketch-based generation and editing results.
    In the top row, given the input text (a) and sketch (b), our method generates realistic 3D results, as shown in (c) and (d) with different views. 
    Users can edit the local regions in a selected view by modifying the text (a) and {sketch (b) in the bottom row}.
    Our method generates high-quality local modification results. 
    }
    \label{fig:results_gen_editing}
\end{figure}

\begin{figure*}[h]
    \centering
    \includegraphics[width=0.97\linewidth]{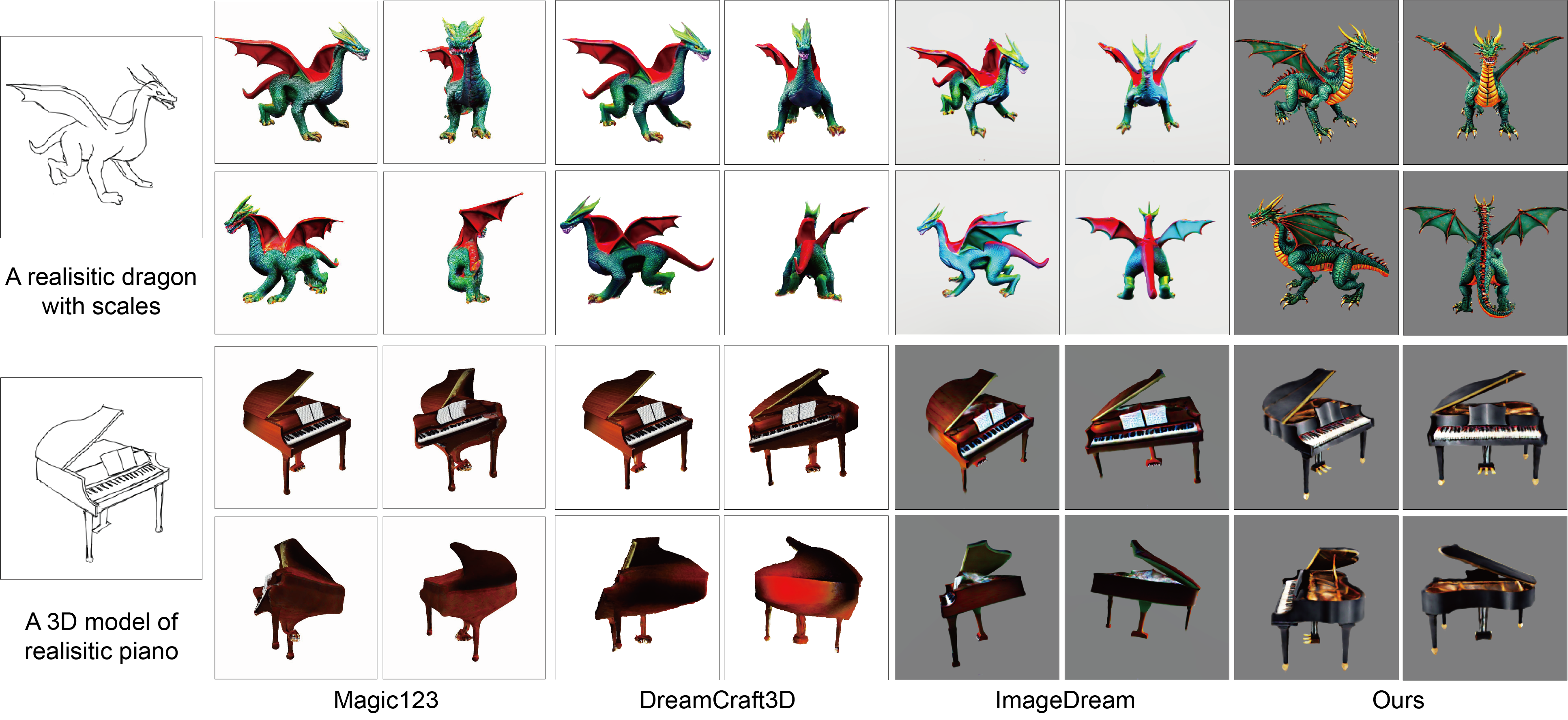}
    \caption{Sketch-based generation comparison. 
    For existing approaches, we first utilize ControlNet \cite{controlnet} to generate 2D images and then use these approaches to generate 3D contents from the 2D images. 
    Magic123 \cite{magic123} generates realistic results in the sketch view but has weird geometry in other views. 
    DreamCraft3D \cite{Dreamcraft3d} generates better results in geometry and texture but still has obvious artifacts. 
    With multi-view information, ImageDream \cite{Imagedream} generates correct geometry but has too light appearance. 
    In contrast, our method generates better results with correct geometry and realistic appearance. 
    }
    \label{fig:generation_compare}
\end{figure*}

\paragraph{Implementation Details}
Our networks are trained and tested on two NVIDIA RTX A6000 GPUs. 
During the training of the depth generation model, the learning rate is 1e-5, the batch size is 64, {and} the number of training steps is 50k. 
To train 3D ControlNet, we set the learning rate to 1e-5, batch size to 4 with gradient accumulation {of} 8, and the number of training steps to 30k.
For the sketch-based generation, we optimize the NeRF representation with 12k steps. 
For the first 10k steps, $\lambda_1=1, \lambda_2=0, \lambda_3=1e2$, and $\lambda_4$ linearly increased from 10.0 to 1000.0 with 20k to 50k steps. 
For the remaining steps, there is a probability of 0.9 to set $\lambda_1=0, \lambda_2=1$, and otherwise $\lambda_1=1, \lambda_2=0$. 
The resolution of images is set to 64 $\times$ 64 for the first 50k steps and changed into 256 $\times$ 256 afterward.
For the sketch-based editing, in the coarse stage, we optimize the NeRF by 50k steps, with $\alpha_1=1.0, \alpha_2=1e5, \alpha_3=1e2$, and $\lambda_4$ {being} linearly increased from 10.0 to 1000.0 with 10k to 30k steps.  
In the fine stage, the hyper-parameters are the same as those for generation, and $\beta_3=1e5$ for the first 10k steps and $\beta_3=1e6$ for the remaining steps. 
We implement sketch-based generation and editing on a single NVIDIA RTX A100. 
It takes 1.0-1.3 hour{s} to generate a single example depending on the area size of the objects in sketches.

\subsection{Results}\label{sec:results}
Our method supports high-quality 3D generation from sketch and text inputs. 
As shown in Fig. \ref{fig:geneation_results}, given text prompts and hand-drawn sketches, our approach generates high-quality 3D results respecting the input sketches and texts. 
Thanks to our sketch-based multi-view generation model, the generated results not only have good quality in the input views of sketches but also have abundant and realistic details in the back view, as shown in the jacket and dolls. 
Our generated results are in large scopes controlled by texts and not restricted to limited categories. 
As shown in Fig. \ref{fig:different_text}, given the same hand-drawn sketches, users can input different text prompts to generate diverse and realistic results. 
It can be seen that sketches and texts provide complementary information: sketches control the geometry of results, while texts control their appearance. 

Our method further supports sketch-based local region editing.
As shown in Fig. \ref{fig:results_editing}, given the NeRFs reconstructed from the real models, users can select arbitrary views and then modify the rendered sketches and provide text descriptions. 
Local editing regions are labeled by 2D masks. 
Our method generates 
realistic editing results with good quality in local editing regions, natural interactions with the original objects (e.g., the new dress and opening treasure chest), {and} unedited regions well preserved. 
Utilizing our method, users can modify the existing components (Fig. \ref{fig:results_editing} (Left)) or add new components (Fig. \ref{fig:results_editing} (Right)). 
Additionally, our method supports further editing of generated 3D contents. 
Fig. \ref{fig:results_gen_editing} (Top) shows a generated result given an input sketch and text prompt.
Then, users can select a view and modify the rendered sketch and text prompt to achieve fine-grained editing of the results.  

\begin{figure*}[h]
    \centering
    \includegraphics[width=0.97\linewidth]{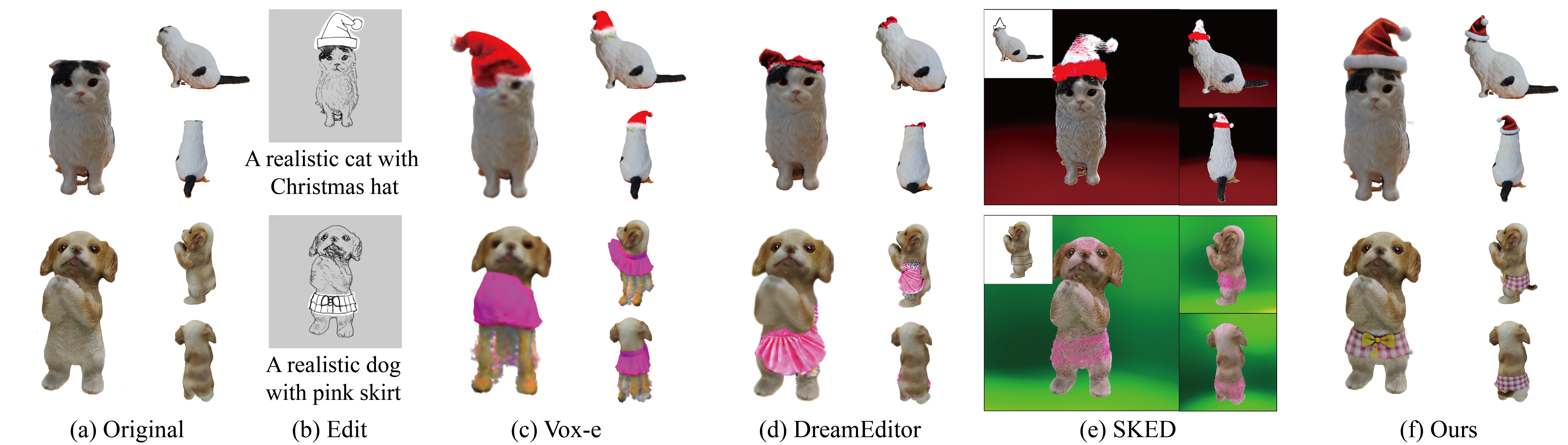}
    \caption{Sketch-based editing comparison. Users can edit the reconstructed real 3D models by changing the texts and editing the rendered sketches. 
    Vox-E \cite{vox-e} and DreamEditor \cite{Dreameditor} only support content editing via text prompts and thus cannot control the shape and editing regions. 
    SKED \cite{Sked} allows editing of local regions with sketches in additional views (shown in the top-left corner), achieving detailed control, but its results are less realistic than ours.
    }
    \label{fig:editing_compare}
\end{figure*}

\subsection{Comparison}\label{sec:Comparison}

\paragraph{Sketch-based Generation.} 
Our method generates realistic 3D contents {given} single-view sketches. 
Since there are no existing approaches to synthesize high-quality results based on sketches and texts directly, we compared our method with an intuitive baseline: 
first utilizing a 2D sketch-to-image generation approach ControlNet \cite{controlnet} to synthesize 2D images and then utilizing state-of-the-art image-to-3D approaches to generate 3D contents.  
We compared our approach with Magic123 \cite{magic123}, DreamCraft3D \cite{Dreamcraft3d}, and ImageDream \cite{Imagedream}, which take single-view images and texts as input for 3D content generation.
As shown in Fig. \ref{fig:generation_compare}, Magic123 generates realistic results in the same views as the input sketches because of the high-quality generation results of 2D ControlNet and faithful input view reconstruction. 
However, the results exhibit weird geometry in other views, such as missing wings in the dragon example and distorted legs in the piano example. 
Utilizing a better pre-trained model (i.e., stable-zero123) and DreamBooth for texture enhancement, DreamCraft3D generates more reasonable 3D geometry and more realistic textures. Still, it leads to obvious artifacts, such as the missing wings and blurry details. 
ImageDream utilizes a pre-trained multi-view diffusion model and solves the structure errors but still has an oblique problem, as shown in the piano example. 
In contrast, our method does not rely on intermediate images and directly generates 3D models from sketches. 
Due to the depth information analysis and multi-view 3D constraint, our method generates more realistic results in terms of both geometry and appearance.

\begin{table}[!t]
    \caption{The quantitative comparisons with sketch-based generation methods, including Magic123 \cite{magic123}, DreamCraft3D \cite{Dreamcraft3d}, and ImageDream \cite{Imagedream}. The abbreviations ``TF'', ``SF'', ``GQ'', and ``TQ'' mean text faithfulness, sketch faithfulness, geometry quality, and texture quality, respectively. The methods are evaluated in terms of the mean value of the CLIP score (CLIP space cosine similarity × 100) and those metrics in the user study. We further report the standard deviation for {the} CLIP score. For all the metrics except \textit{{STD}}, the higher, the better.}
    \label{tab:text_generation}
    \begin{center}
    \resizebox{\columnwidth}{!}{
    \begin{tabular}{lcc|cccc|c}
        \toprule
         & \multicolumn{2}{c}{\textit{CLIP}} & \multicolumn{5}{c}{\textit{Human Preference}}\\
        \textit{Method} & \textit{Mean} & \textit{{STD}} & TF & SF & GQ & TQ & Overall \\\midrule
        Magic123 \quad
        & 27.61 & 3.18 & 3.18 & 3.19 & 2.78 & 2.86 & 2.76 \\
        DreamCraft3D \quad
        & 29.34 & 2.56 & 3.35 & 3.32 & 2.95 & 2.91 & 2.87 \\
        ImageDream \quad
        & 27.61 & 2.75 & 3.40 & 3.49 & 3.06 & 2.89 & 2.97 \\
        Ours  & \textbf{30.90} & \textbf{2.13} & \textbf{4.71} & \textbf{4.52} & \textbf{4.67} & \textbf{4.72} & \textbf{4.72} \\
    \end{tabular}}
    \end{center}
\end{table}

\begin{table}[!t]
    \caption{The quantitative comparisons with {SKED}~\cite{Sked} for sketch-based editing.
    The abbreviations ``TF'', ``SF'', ``PU'', and ``EQ'' mean text faithfulness, sketch faithfulness, preservation of unedited regions, and editing component quality, respectively. The methods are evaluated in terms of the mean value of {the} CLIP score and those metrics in the user study. The standard deviation for {the} CLIP score {is} also included.}
    \label{tab:text_editing}
    \begin{center}
    \resizebox{\columnwidth}{!}{
    \begin{tabular}{lcc|cccc|c}
        \toprule
         & \multicolumn{2}{c}{\textit{CLIP}} & \multicolumn{5}{c}{\textit{Human Preference}}\\
        \textit{Method} & \textit{Mean} & \textit{STD} & TF & SF & PU & EQ & Overall \\\midrule
        SKED 
        \qquad \qquad & 32.69 & 3.05  & 2.46 & 2.50 & 3.02 & 2.24 & 2.33 \\
        Ours  & \textbf{34.16} & \textbf{2.50} & \textbf{4.86} & \textbf{4.86} & \textbf{4.84} & \textbf{4.82} & \textbf{4.84} \\
    \end{tabular}}
    \end{center}
\end{table}

\paragraph{Sketch-based Editing.}
Our method supports local editing of real objects based on sketches and text. 
To show its advantages, we compare it with existing local editing approaches, including Vox-E \cite{vox-e}, DreamEditor \cite{Dreameditor}, and SKED \cite{Sked}.
Vox-E utilizes a voxel grid to reconstruct real objects and achieves text-based local editing. 
However, since the editing region has to be predicted by texts, it does not support accurate control of editing regions and results. For example, the left eye of the cat is undesirably changed, and the shape and texture patterns of the dog example are unexpected.
Dream{E}ditor \cite{Dreameditor} also utilizes text to predict editing regions {and} thus might have mistaken editing regions such as the back view of dogs. Additionally, due to the NeRF definition on original meshes, it cannot support large-scale geometry editing, such as adding a Christmas hat.
To achieve a more detailed control, SKED takes multi-view sketches with texts as input. However, it generates blurry results in the edited regions and slightly changes the unedited regions. 
Additionally, since SKED translates the input sketches into binary masks, it cannot control the texture details like the wrinkles of the hat and the decorative patterns of the skirt.  
In contrast, our method generates the most realistic editing results. 

\paragraph{Quantitative Study.} We assess our method and existing baseline approaches using CLIP Score~\cite{CLIP}. This metric evaluates the correlation between the text prompts and rendering images by embedding them into a shared latent space and calculating the cosine similarity. The reasonableness of this metric is discussed in the supplemental material. For sketch-based generation, we compare our method against existing approaches on 20 examples. These examples involve a wide set of categories, ranging from animals, architecture, food, vehicles, and instruments.
We personally crafted the prompts for these examples, providing concise descriptions of simple, everyday objects. 
As shown in Table~\ref{tab:text_generation}, compared with the combination of 2D ControlNet and Image-to-3D baseline approaches, our method achieves the highest scores, thus validating the better alignment of our results with text prompts. 
For editing, we compare with the sketch- and text-based editing approach, SKED, on 15 editing examples. 
Similarly, these examples {exhibit} good diversity involving different categories.
As shown in Table~\ref{tab:text_editing}, our method also has the highest scores, validating the better editing alignment for text prompts. 
The examples used to calculate the CLIP scores are given in the supplement materials.

\begin{figure}[t]
\centering
\begin{tabular}{cc}
\hspace{-3mm}
\rotatebox{90}{\scriptsize{\hspace{3mm}(a) Input}}
&
\hspace{-3mm}
\includegraphics[width=0.6\linewidth]{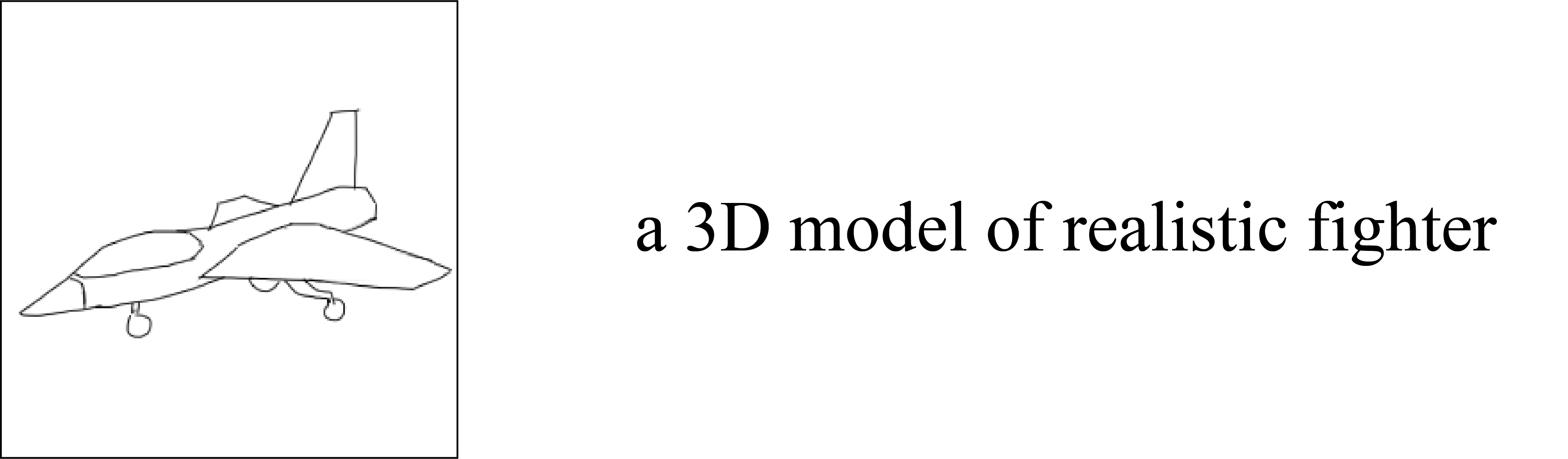} \\

\hspace{-3mm}
\rotatebox{90}{\scriptsize{\hspace{2mm}(b) w/o Depth}}
&
\hspace{-3mm}
\includegraphics[width=0.92\linewidth]{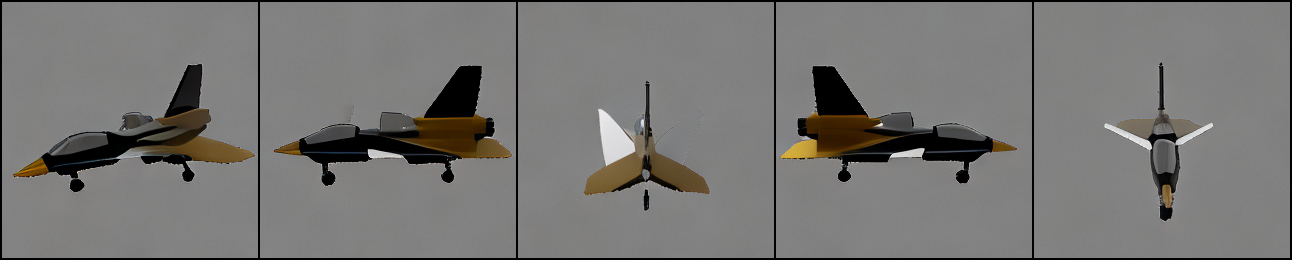} \\

\hspace{-3mm}
\rotatebox{90}{\scriptsize{\hspace{1mm}(c) w/o 3D Atten}}
&
\hspace{-3mm}
\includegraphics[width=0.92\linewidth]{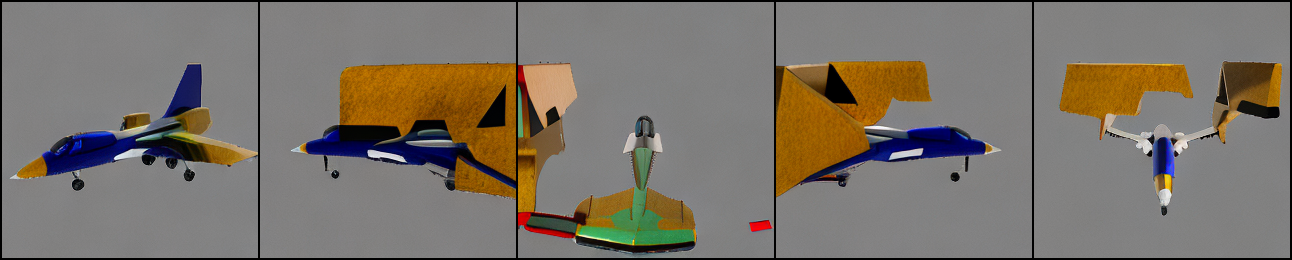} \\

\hspace{-3mm}
\rotatebox{90}{\scriptsize{\hspace{3mm}(d) w/o Near}}
&
\hspace{-3mm}
\includegraphics[width=0.92\linewidth]{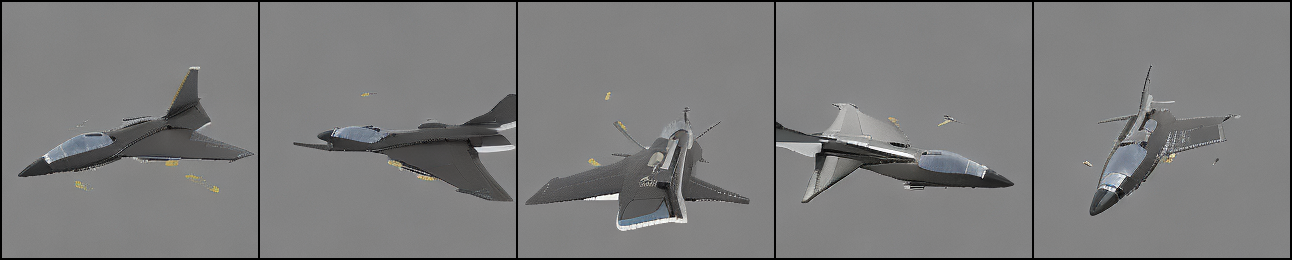}\\

\hspace{-3mm}
{\rotatebox{90}{\scriptsize{\hspace{4mm}(e) Ours}}}
&
\hspace{-3mm}
\includegraphics[width=0.92\linewidth]{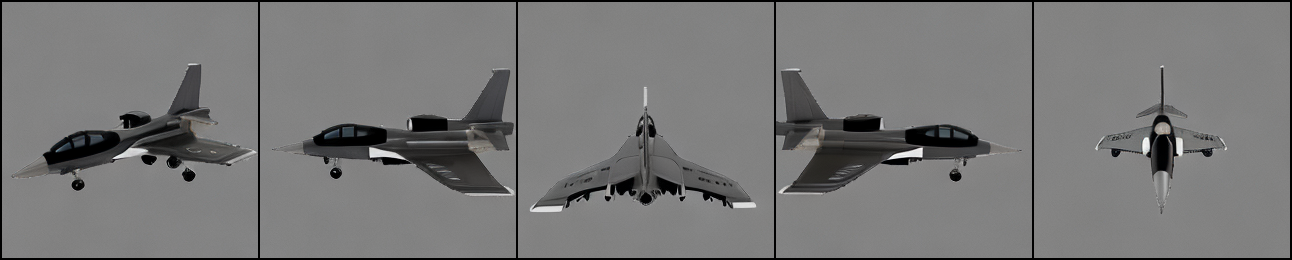}\\
        
\end{tabular}
\caption{Ablation study of the Multi-View ControlNet. 
Given a hand-drawn sketch and a text prompt (a), without the depth-guided warning (b), the generated results have good quality in the sketch view but are consistent across views.
Without the 3D attention module (c), the results have strange components in the views other than the sketch view.
If the input sketch is warped into all views (d), the generated images might suffer from twisted geometry. 
Our full method (e) generates the most realistic results in the input and novel views with good 3D consistency. 
}
\label{fig:ablaton_sampling}
\end{figure}

\begin{figure*}[h]
    \centering
    \includegraphics[width=0.97\linewidth]{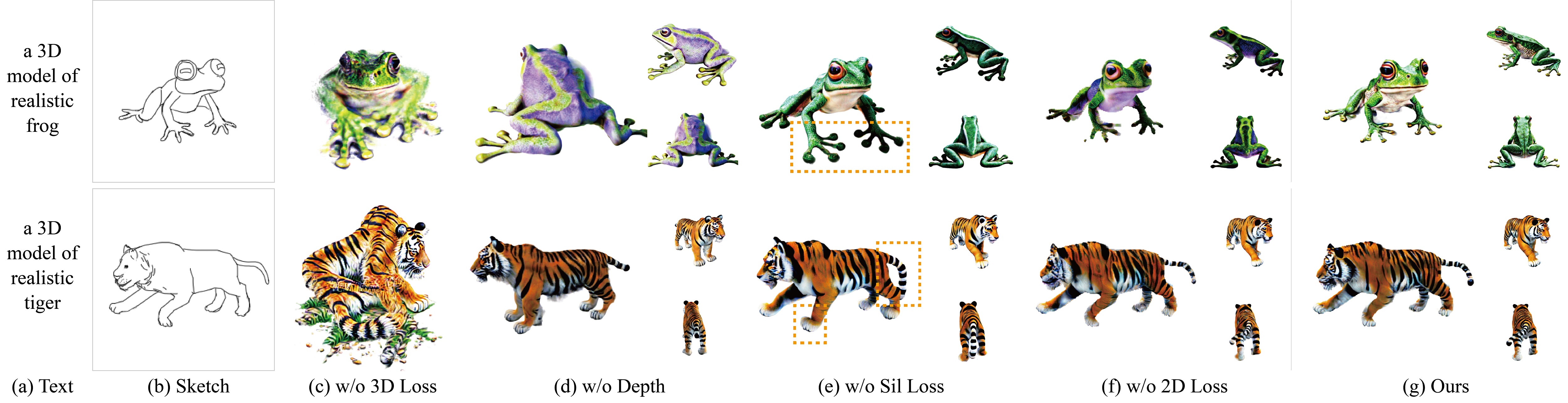}
    \caption{Ablation study of sketch-based generation. 
    The results without the 3D loss (c) are realistic but have a low correlation with the input sketches (b).
    Without the depth-guided warping (d), the generated results are low in faithfulness to the input sketches.
    Without the silhouette loss (e), the generated results misalign with sketches in local details, such as the feet and tail.
    Without the 2D loss (f), the results tend to be fuzzy.}
    \label{fig:ablation_generation}
\end{figure*}

\subsection{Ablation Study}\label{sec:ablation}
We conduct ablation studies to prove the effectiveness of the key components in our framework. 
Specifically, we disable the key components of the sketch-based multi-view diffusion model to show their impacts. 
Then, we show the effectiveness of each loss term for the sketch-based 3D generation. 
For sketch-based editing, we also remove the two-stage strategy and local enhancement to prove their necessity.

In the sketch-based multi-view image generation model, we predict depth maps to warp the input sketches, explicitly building the spatial correspondence. 
As shown in Fig. \ref{fig:ablaton_sampling} (b), without the depth warp strategy, the generated results have low 3D consistency, leading to different colors in the aircraft wings and additional parts in the back view. 
In order to add sketch control into the pre-trained MVDream, we construct ControlNet with 3D attention. 
If we remove it and feed conditional inputs independently, the generated results tend to have strange parts in novel views due to the lack of 3D information. 
Since the predicted depth maps have slight errors, we only warp the input sketch into the nearest view, whose results are most reliable. If we warp it into all views, the warped sketches in far views have low quality, which largely affects the generation results, as shown in row (d). 
In contrast, our full model generates the best results with realistic appearance and good 3D consistency. 

For sketch-based 3D generation, to add a sketch constraint in novel views, we utilize the SDS loss based on the sketch-based multi-view diffusion model. 
As shown in Fig. \ref{fig:ablation_generation} (c), if we remove the 3D Loss, the generated results show good realism but have no relationship with the input sketches. 
We utilize the depth warp strategy to connect novel views with the input sketch view explicitly. 
As shown in Fig. \ref{fig:ablation_generation} (d), without this strategy, sketches sometimes cannot control the viewpoints of the generated 3D contents. 
The control of details is also affected, e.g., causing an undesired tiger pose. 
During optimization, we utilize a silhouette loss to improve the detail control further. 
Without this silhouette loss (Fig.\ref{fig:ablation_generation} (e)), local details tend to misalign with sketches, such as the bigger frog feet and a tiger tail with a different shape. 
Additionally, if we remove the 2D loss during optimization, the generated results (Fig.\ref{fig:ablation_generation} (f)) have more fuzzy details, such as the dim frog texture and blurry tiger hair. 
Our full approach generates the most realistic results with fine details and the most faithfulness to the input sketches. 

\begin{figure*}[h]
    \centering
    \includegraphics[width=1.0\linewidth]{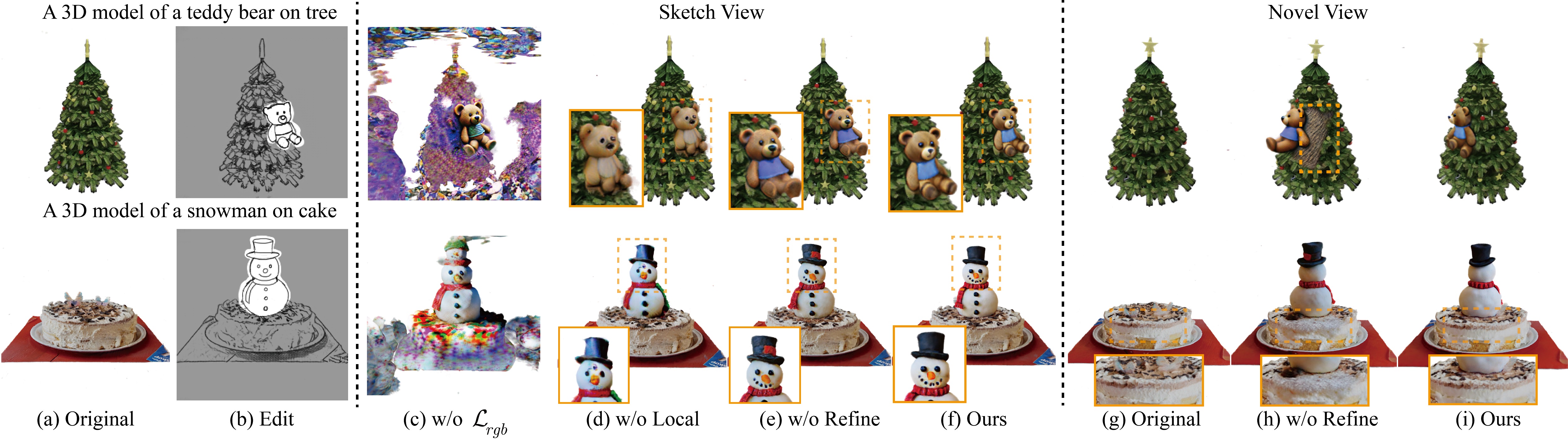}
    \caption{Ablation study of sketch-based editing.
    The original 3D models (a) and the editing texts and sketches (b) are shown in the left regions. Without the image loss (c), unedited regions are totally changed. Without the local enhancement (d), the generated components are fuzzy and have obvious artifacts like the strange eyes. Replacing the fine stage mask with {the} coarse mask, the editing results are acceptable in view of sketches (e), but the unedited regions are changed in novel views (h).  
    }
    \label{fig:ablation_editing}
\end{figure*}
 
For sketch-based editing, the unedited regions should be retained while the edited components should be effectively modified. 
As shown in Fig. \ref{fig:ablation_editing} (c), if the image loss with the original objects is removed, the whole objects are totally changed and have obvious floaters due to the lack of constraint with the original objects. 
When the local enhancement strategy is removed, the quality of edited regions is degraded. 
The faithfulness with the sketch is also affected. See the missing clothes of the bear and the smiles of the snowman in Fig. \ref{fig:ablation_editing} (d). 
Our editing process includes two-stage optimization. 
As shown in (e), directly using the coarse 3D masks without the refined 3D masks has limited influences in the sketch view. 
However, in a novel view, unedited regions in the generated results (h) are undesirably modified, such as the strange bark and new cake pattern, which are included in coarse 3D masks. 
Compared with all the baseline approaches, our full method generates the most realistic results on the editing components while preserving the unedited regions.

\subsection{Perception Study}\label{sec:user_study}
We conducted a perception study to validate our method using human eyes. For sketch-based generation, we compare with the same set of approaches in qualitative comparison. The same 20 cases are also utilized, each containing an input sketch, a text prompt, and the synthesized 3D contents. 
Users evaluated the results on a scale of 1-5, where `1' means very poor, while `5' indicates excellent, across five aspects: text faithfulness, sketch faithfulness, geometry quality, texture quality, and overall quality.  
All methods presented video results of generated models, showcasing a full rotation in azimuth and a fixed elevation of 15 degrees. For each participant, we randomly selected 5 cases from the entire set, resulting in $ 5 \times 5 = 25$ answers per user. 
In total, 41 participants aged 18-40 without professional drawing skills contributed, yielding a dataset of $ 25 \times 41 = 1025$ answers. Our method outperformed existing baseline approaches, achieving the highest scores in all aspects according to Table \ref{tab:text_generation}. Notably, our approach excelled in text faithfulness but showed a slight weakness in sketch faithfulness. In the supplementary materials, box plots are drawn to show the evaluation scores. We also report the one-way ANOVA tests and paired T-tests to validate the superior performance of our approach.

For sketch-based editing, we employed a similar setup to the generation study but compare{d} with SKED. Users also evaluated the results using a 1-5 scale across five aspects: text faithfulness, sketch faithfulness, preservation of unedited regions, editing component quality, and overall quality. For each user, we collected $ 5 \times 2 = 10$ answers, resulting in a total of $10 \times 41 = 410$ answers. 
As shown in Table \ref{tab:text_editing}, our method achieved better results compared with {the} baseline approach in all aspects. Similar to the generation study, our approach excelled in text faithfulness. However, it showed a slight weakness in the quality of newly generated components, a key challenge in the editing process.
Similar to sketch-based generation, we further report the box plot, one-way ANOVA tests, and paired T-test in the supplementary materials.

%% file: sections/6_conclusion.tex
\section{Conclusion}

In this paper, we have presented a sketch-based 3D generation and editing method to create realistic high-quality 3D models {from the input text prompts} and hand-drawn sketches. 
In order to supplement the missing appearance and propagate the single-view sketch into 3D space, we design a sketch-based multi-view image generation diffusion model. 
To improve the sketch faithfulness and generation quality, we translate input sketches into depth maps, which are utilized to warp sketches into novel views to build 3D correspondence. 
The original and warped sketches serve as input conditions of 3D ControlNet, which has a 3D attention module to generate 3D consistent multi-view images. 
To generate realistic 3D contents based on sketches, we utilize the 3D SDS of sketch-based multi-view diffusion and 2D ISM of text-to-image diffusion, optimizing high-quality NeRF models.
We further propose a sketch-based local editing method, which includes a coarse stage to get a precise editing 3D mask, and a fine stage with local enhancement to generate high-quality editing results. 
Extensive experiments show the advantages of our method over baseline approaches.

\begin{figure}[t]
    \centering
    \includegraphics[width=1.0\linewidth]{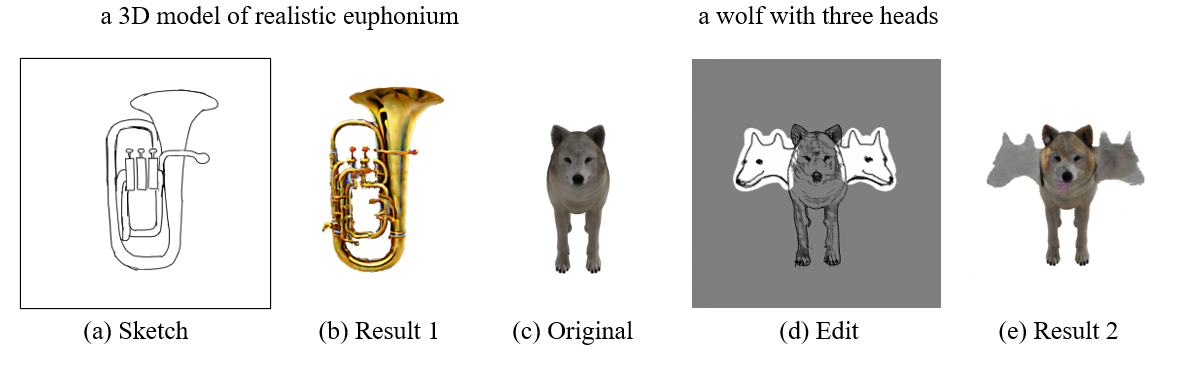}
    \caption{The failure cases of our method. For sketch-based generation, given sketches (a) and text (above the images), our method generates results (b) that are misaligned with the sketches in detail. 
    Additionally, our method cannot edit examples that very rare in dataset. 
    }
    \label{fig:failure_cases}
\end{figure}

Thanks to the good generalization of the pre-trained diffusion model, our method produces generated results that are not constrained to limited categories. However, as illustrated in Fig. \ref{fig:failure_cases}, challenges arise for cases that are rare in the training dataset. In such instances, while our method generates a coarse result, the sketch faithfulness and realism may be influenced.
Similarly, for sketch-based editing, our method faces limitations in handling what can be deemed 'too strange cases,' often resulting in degraded editing results with fuzzy details.
For future work, although our method supports the appearance control based on text prompts, it is still difficult for users to achieve detail control of lighting, color, and material. 
Additionally, hand-drawn sketches loosely control the shape and layout while lacking too detailed control like the mushroom dots in Fig. \ref{fig:teaser}.
Adding additional conditions, such as color strokes, can partly solve this problem. 
Besides, our current system currently requires about 1 hour for generation and 1.5 hours for editing, and thus does not support interactive generation or editing.
A potential solution to accelerate the sketch-based generation and editing process is to design 3D native generation pipelines \cite{LRM} trained by sketch constraints.

\begin{acks} 
This work was supported by National Natural Science Foundation of China (No. 62322210), Beijing Municipal Natural Science Foundation for Distinguished Young Scholars (No. JQ21013), and Beijing Municipal Science and Technology Commission (No. Z231100005923031). 
\end{acks}

%% file: sample-acmtog-SIGGRAPH-submission.bbl

\begin{thebibliography}{74}


\ifx \showCODEN    \undefined \def \showCODEN     #1{\unskip}     \fi
\ifx \showDOI      \undefined \def \showDOI       #1{#1}\fi
\ifx \showISBNx    \undefined \def \showISBNx     #1{\unskip}     \fi
\ifx \showISBNxiii \undefined \def \showISBNxiii  #1{\unskip}     \fi
\ifx \showISSN     \undefined \def \showISSN      #1{\unskip}     \fi
\ifx \showLCCN     \undefined \def \showLCCN      #1{\unskip}     \fi
\ifx \shownote     \undefined \def \shownote      #1{#1}          \fi
\ifx \showarticletitle \undefined \def \showarticletitle #1{#1}   \fi
\ifx \showURL      \undefined \def \showURL       {\relax}        \fi
\providecommand\bibfield[2]{#2}
\providecommand\bibinfo[2]{#2}
\providecommand\natexlab[1]{#1}
\providecommand\showeprint[2][]{arXiv:#2}

\bibitem[Chan et~al\mbox{.}(2022)]%
        {Line_drawing}
\bibfield{author}{\bibinfo{person}{Caroline Chan}, \bibinfo{person}{Fr{\'e}do Durand}, {and} \bibinfo{person}{Phillip Isola}.} \bibinfo{year}{2022}\natexlab{}.
\newblock \showarticletitle{Learning to generate line drawings that convey geometry and semantics}. In \bibinfo{booktitle}{\emph{Proceedings of the IEEE/CVF Conference on Computer Vision and Pattern Recognition}}. \bibinfo{pages}{7915--7925}.
\newblock


\bibitem[Chen et~al\mbox{.}(2003)]%
        {2003visual_retrieval}
\bibfield{author}{\bibinfo{person}{Ding-Yun Chen}, \bibinfo{person}{Xiao-Pei Tian}, \bibinfo{person}{Yu-Te Shen}, {and} \bibinfo{person}{Ming Ouhyoung}.} \bibinfo{year}{2003}\natexlab{}.
\newblock \showarticletitle{On visual similarity based {3D} model retrieval}. In \bibinfo{booktitle}{\emph{Computer Graphics Forum}}, Vol.~\bibinfo{volume}{22}. Wiley Online Library, \bibinfo{pages}{223--232}.
\newblock


\bibitem[Chen et~al\mbox{.}(2023a)]%
        {Fantasia3D}
\bibfield{author}{\bibinfo{person}{Rui Chen}, \bibinfo{person}{Yongwei Chen}, \bibinfo{person}{Ningxin Jiao}, {and} \bibinfo{person}{Kui Jia}.} \bibinfo{year}{2023}\natexlab{a}.
\newblock \showarticletitle{Fantasia3D: Disentangling Geometry and Appearance for High-quality Text-to-3D Content Creation}. In \bibinfo{booktitle}{\emph{Proceedings of the IEEE/CVF International Conference on Computer Vision}}. \bibinfo{pages}{22189--22199}.
\newblock


\bibitem[Chen et~al\mbox{.}(2009)]%
        {Sketch2photo}
\bibfield{author}{\bibinfo{person}{Tao Chen}, \bibinfo{person}{Ming{-}Ming Cheng}, \bibinfo{person}{Ping Tan}, \bibinfo{person}{Ariel Shamir}, {and} \bibinfo{person}{Shi{-}Min Hu}.} \bibinfo{year}{2009}\natexlab{}.
\newblock \showarticletitle{Sketch2Photo: internet image montage}.
\newblock \bibinfo{journal}{\emph{ACM Transactions on Graphics}} \bibinfo{volume}{28}, \bibinfo{number}{5} (\bibinfo{year}{2009}), \bibinfo{pages}{124}.
\newblock


\bibitem[Chen et~al\mbox{.}(2023b)]%
        {control3d}
\bibfield{author}{\bibinfo{person}{Yang Chen}, \bibinfo{person}{Yingwei Pan}, \bibinfo{person}{Yehao Li}, \bibinfo{person}{Ting Yao}, {and} \bibinfo{person}{Tao Mei}.} \bibinfo{year}{2023}\natexlab{b}.
\newblock \showarticletitle{Control3d: Towards controllable text-to-3d generation}. In \bibinfo{booktitle}{\emph{Proceedings of the 31st ACM International Conference on Multimedia}}. \bibinfo{pages}{1148--1156}.
\newblock


\bibitem[Cheng et~al\mbox{.}(2023)]%
        {progressive3d}
\bibfield{author}{\bibinfo{person}{Xinhua Cheng}, \bibinfo{person}{Tianyu Yang}, \bibinfo{person}{Jianan Wang}, \bibinfo{person}{Yu Li}, \bibinfo{person}{Lei Zhang}, \bibinfo{person}{Jian Zhang}, {and} \bibinfo{person}{Li Yuan}.} \bibinfo{year}{2023}\natexlab{}.
\newblock \showarticletitle{{Progressive3D}: Progressively local editing for text-to-{3D} content creation with complex semantic prompts}.
\newblock \bibinfo{journal}{\emph{arXiv preprint arXiv:2310.11784}} (\bibinfo{year}{2023}).
\newblock


\bibitem[Deitke et~al\mbox{.}(2023)]%
        {objaverse}
\bibfield{author}{\bibinfo{person}{Matt Deitke}, \bibinfo{person}{Dustin Schwenk}, \bibinfo{person}{Jordi Salvador}, \bibinfo{person}{Luca Weihs}, \bibinfo{person}{Oscar Michel}, \bibinfo{person}{Eli VanderBilt}, \bibinfo{person}{Ludwig Schmidt}, \bibinfo{person}{Kiana Ehsani}, \bibinfo{person}{Aniruddha Kembhavi}, {and} \bibinfo{person}{Ali Farhadi}.} \bibinfo{year}{2023}\natexlab{}.
\newblock \showarticletitle{Objaverse: A universe of annotated {3D} objects}. In \bibinfo{booktitle}{\emph{Proceedings of the IEEE/CVF Conference on Computer Vision and Pattern Recognition}}. \bibinfo{pages}{13142--13153}.
\newblock


\bibitem[Delanoy et~al\mbox{.}(2018)]%
        {delanoy2018_3d_volumn}
\bibfield{author}{\bibinfo{person}{Johanna Delanoy}, \bibinfo{person}{Mathieu Aubry}, \bibinfo{person}{Phillip Isola}, \bibinfo{person}{Alexei~A Efros}, {and} \bibinfo{person}{Adrien Bousseau}.} \bibinfo{year}{2018}\natexlab{}.
\newblock \showarticletitle{{3D} sketching using multi-view deep volumetric prediction}.
\newblock \bibinfo{journal}{\emph{Proceedings of the ACM on Computer Graphics and Interactive Techniques}} \bibinfo{volume}{1}, \bibinfo{number}{1} (\bibinfo{year}{2018}), \bibinfo{pages}{1--22}.
\newblock


\bibitem[Fehn(2004)]%
        {fehn2004depth}
\bibfield{author}{\bibinfo{person}{Christoph Fehn}.} \bibinfo{year}{2004}\natexlab{}.
\newblock \showarticletitle{Depth-image-based rendering (DIBR), compression, and transmission for a new approach on 3D-TV}. In \bibinfo{booktitle}{\emph{Stereoscopic displays and virtual reality systems XI}}, Vol.~\bibinfo{volume}{5291}. \bibinfo{pages}{93--104}.
\newblock


\bibitem[Funkhouser et~al\mbox{.}(2003)]%
        {2003_search_sketch_generation}
\bibfield{author}{\bibinfo{person}{Thomas Funkhouser}, \bibinfo{person}{Patrick Min}, \bibinfo{person}{Michael Kazhdan}, \bibinfo{person}{Joyce Chen}, \bibinfo{person}{Alex Halderman}, \bibinfo{person}{David Dobkin}, {and} \bibinfo{person}{David Jacobs}.} \bibinfo{year}{2003}\natexlab{}.
\newblock \showarticletitle{A search engine for {3D} models}.
\newblock \bibinfo{journal}{\emph{ACM Transactions on Graphics}} \bibinfo{volume}{22}, \bibinfo{number}{1} (\bibinfo{year}{2003}), \bibinfo{pages}{83--105}.
\newblock


\bibitem[Gao et~al\mbox{.}(2022)]%
        {gao2022_sketchsampler}
\bibfield{author}{\bibinfo{person}{Chenjian Gao}, \bibinfo{person}{Qian Yu}, \bibinfo{person}{Lu Sheng}, \bibinfo{person}{Yi-Zhe Song}, {and} \bibinfo{person}{Dong Xu}.} \bibinfo{year}{2022}\natexlab{}.
\newblock \showarticletitle{{SketchSampler}: Sketch-Based {3D} Reconstruction via View-Dependent Depth Sampling}. In \bibinfo{booktitle}{\emph{European Conference on Computer Vision}}. \bibinfo{pages}{464--479}.
\newblock


\bibitem[Gao et~al\mbox{.}(2023b)]%
        {SketchFaceNeRF}
\bibfield{author}{\bibinfo{person}{Lin Gao}, \bibinfo{person}{Feng{-}Lin Liu}, \bibinfo{person}{Shu{-}Yu Chen}, \bibinfo{person}{Kaiwen Jiang}, \bibinfo{person}{Chun{-}Peng Li}, \bibinfo{person}{Yu{-}Kun Lai}, {and} \bibinfo{person}{Hongbo Fu}.} \bibinfo{year}{2023}\natexlab{b}.
\newblock \showarticletitle{{SketchFaceNeRF}: Sketch-based Facial Generation and Editing in Neural Radiance Fields}.
\newblock \bibinfo{journal}{\emph{ACM Transactions on Graphics}} \bibinfo{volume}{42}, \bibinfo{number}{4} (\bibinfo{year}{2023}), \bibinfo{pages}{159:1--159:17}.
\newblock


\bibitem[Gao et~al\mbox{.}(2023a)]%
        {TextDeformer}
\bibfield{author}{\bibinfo{person}{William Gao}, \bibinfo{person}{Noam Aigerman}, \bibinfo{person}{Thibault Groueix}, \bibinfo{person}{Vova Kim}, {and} \bibinfo{person}{Rana Hanocka}.} \bibinfo{year}{2023}\natexlab{a}.
\newblock \showarticletitle{{TextDeformer}: Geometry Manipulation Using Text Guidance}. In \bibinfo{booktitle}{\emph{ACM SIGGRAPH 2023 Conference Proceedings}}. \bibinfo{pages}{82:1--82:11}.
\newblock


\bibitem[Garbin et~al\mbox{.}(2022)]%
        {VolTeMorph}
\bibfield{author}{\bibinfo{person}{Stephan~J. Garbin}, \bibinfo{person}{Marek Kowalski}, \bibinfo{person}{Virginia Estellers}, \bibinfo{person}{Stanislaw Szymanowicz}, \bibinfo{person}{Shideh Rezaeifar}, \bibinfo{person}{Jingjing Shen}, \bibinfo{person}{Matthew Johnson}, {and} \bibinfo{person}{Julien Valentin}.} \bibinfo{year}{2022}\natexlab{}.
\newblock \showarticletitle{VolTeMorph: Realtime, Controllable and Generalisable Animation of Volumetric Representations}.
\newblock \bibinfo{journal}{\emph{CoRR}}  \bibinfo{volume}{abs/2208.00949} (\bibinfo{year}{2022}).
\newblock


\bibitem[Hong et~al\mbox{.}(2023)]%
        {LRM}
\bibfield{author}{\bibinfo{person}{Yicong Hong}, \bibinfo{person}{Kai Zhang}, \bibinfo{person}{Jiuxiang Gu}, \bibinfo{person}{Sai Bi}, \bibinfo{person}{Yang Zhou}, \bibinfo{person}{Difan Liu}, \bibinfo{person}{Feng Liu}, \bibinfo{person}{Kalyan Sunkavalli}, \bibinfo{person}{Trung Bui}, {and} \bibinfo{person}{Hao Tan}.} \bibinfo{year}{2023}\natexlab{}.
\newblock \showarticletitle{{LRM:} Large Reconstruction Model for Single Image to 3D}.
\newblock \bibinfo{journal}{\emph{CoRR}}  \bibinfo{volume}{abs/2311.04400} (\bibinfo{year}{2023}).
\newblock


\bibitem[Igarashi et~al\mbox{.}(2006)]%
        {igarashi2006teddy}
\bibfield{author}{\bibinfo{person}{Takeo Igarashi}, \bibinfo{person}{Satoshi Matsuoka}, {and} \bibinfo{person}{Hidehiko Tanaka}.} \bibinfo{year}{2006}\natexlab{}.
\newblock \showarticletitle{Teddy: a sketching interface for {3D} freeform design}.
\newblock In \bibinfo{booktitle}{\emph{ACM SIGGRAPH 2006 Courses}}. \bibinfo{pages}{11--es}.
\newblock


\bibitem[Isola et~al\mbox{.}(2017)]%
        {pix2pix}
\bibfield{author}{\bibinfo{person}{Phillip Isola}, \bibinfo{person}{Jun{-}Yan Zhu}, \bibinfo{person}{Tinghui Zhou}, {and} \bibinfo{person}{Alexei~A. Efros}.} \bibinfo{year}{2017}\natexlab{}.
\newblock \showarticletitle{Image-to-Image Translation with Conditional Adversarial Networks}. In \bibinfo{booktitle}{\emph{Proceedings of the IEEE/CVF Conference on Computer Vision and Pattern Recognition}}. \bibinfo{pages}{5967--5976}.
\newblock


\bibitem[Li et~al\mbox{.}(2020)]%
        {li2020sketch2cad}
\bibfield{author}{\bibinfo{person}{Changjian Li}, \bibinfo{person}{Hao Pan}, \bibinfo{person}{Adrien Bousseau}, {and} \bibinfo{person}{Niloy~J Mitra}.} \bibinfo{year}{2020}\natexlab{}.
\newblock \showarticletitle{{Sketch2CAD}: Sequential {CAD} modeling by sketching in context}.
\newblock \bibinfo{journal}{\emph{ACM Transactions on Graphics}} \bibinfo{volume}{39}, \bibinfo{number}{6} (\bibinfo{year}{2020}), \bibinfo{pages}{164:1--164:14}.
\newblock


\bibitem[Li et~al\mbox{.}(2022)]%
        {li2022free2cad}
\bibfield{author}{\bibinfo{person}{Changjian Li}, \bibinfo{person}{Hao Pan}, \bibinfo{person}{Adrien Bousseau}, {and} \bibinfo{person}{Niloy~J Mitra}.} \bibinfo{year}{2022}\natexlab{}.
\newblock \showarticletitle{{Free2CAD}: Parsing freehand drawings into CAD commands}.
\newblock \bibinfo{journal}{\emph{ACM Transactions on Graphics}} \bibinfo{volume}{41}, \bibinfo{number}{4} (\bibinfo{year}{2022}), \bibinfo{pages}{93:1--93:16}.
\newblock


\bibitem[Li et~al\mbox{.}(2018)]%
        {li2018robust_surface}
\bibfield{author}{\bibinfo{person}{Changjian Li}, \bibinfo{person}{Hao Pan}, \bibinfo{person}{Yang Liu}, \bibinfo{person}{Xin Tong}, \bibinfo{person}{Alla Sheffer}, {and} \bibinfo{person}{Wenping Wang}.} \bibinfo{year}{2018}\natexlab{}.
\newblock \showarticletitle{Robust flow-guided neural prediction for sketch-based freeform surface modeling}.
\newblock \bibinfo{journal}{\emph{ACM Transactions on Graphics}} \bibinfo{volume}{37}, \bibinfo{number}{6} (\bibinfo{year}{2018}), \bibinfo{pages}{238:1--238:12}.
\newblock


\bibitem[Li et~al\mbox{.}(2024)]%
        {focaldreamer}
\bibfield{author}{\bibinfo{person}{Yuhan Li}, \bibinfo{person}{Yishun Dou}, \bibinfo{person}{Yue Shi}, \bibinfo{person}{Yu Lei}, \bibinfo{person}{Xuanhong Chen}, \bibinfo{person}{Yi Zhang}, \bibinfo{person}{Peng Zhou}, {and} \bibinfo{person}{Bingbing Ni}.} \bibinfo{year}{2024}\natexlab{}.
\newblock \showarticletitle{FocalDreamer: Text-Driven 3D Editing via Focal-Fusion Assembly}. In \bibinfo{booktitle}{\emph{Conference on Artificial Intelligence}}. \bibinfo{pages}{3279--3287}.
\newblock


\bibitem[Li et~al\mbox{.}(2023)]%
        {mvcontrol}
\bibfield{author}{\bibinfo{person}{Zhiqi Li}, \bibinfo{person}{Yiming Chen}, \bibinfo{person}{Lingzhe Zhao}, {and} \bibinfo{person}{Peidong Liu}.} \bibinfo{year}{2023}\natexlab{}.
\newblock \showarticletitle{MVControl: Adding Conditional Control to Multi-view Diffusion for Controllable Text-to-3D Generation}.
\newblock \bibinfo{journal}{\emph{arXiv preprint arXiv:2311.14494}} (\bibinfo{year}{2023}).
\newblock


\bibitem[Liang et~al\mbox{.}(2023)]%
        {luciddreamer}
\bibfield{author}{\bibinfo{person}{Yixun Liang}, \bibinfo{person}{Xin Yang}, \bibinfo{person}{Jiantao Lin}, \bibinfo{person}{Haodong Li}, \bibinfo{person}{Xiaogang Xu}, {and} \bibinfo{person}{Yingcong Chen}.} \bibinfo{year}{2023}\natexlab{}.
\newblock \showarticletitle{{LucidDreamer}: Towards High-Fidelity Text-to-{3D} Generation via Interval Score Matching}.
\newblock \bibinfo{journal}{\emph{arXiv preprint arXiv:2311.11284}} (\bibinfo{year}{2023}).
\newblock


\bibitem[Lin et~al\mbox{.}(2023)]%
        {magic3d}
\bibfield{author}{\bibinfo{person}{Chen-Hsuan Lin}, \bibinfo{person}{Jun Gao}, \bibinfo{person}{Luming Tang}, \bibinfo{person}{Towaki Takikawa}, \bibinfo{person}{Xiaohui Zeng}, \bibinfo{person}{Xun Huang}, \bibinfo{person}{Karsten Kreis}, \bibinfo{person}{Sanja Fidler}, \bibinfo{person}{Ming-Yu Liu}, {and} \bibinfo{person}{Tsung-Yi Lin}.} \bibinfo{year}{2023}\natexlab{}.
\newblock \showarticletitle{{Magic3D}: High-resolution text-to-{3D} content creation}. In \bibinfo{booktitle}{\emph{Proceedings of the IEEE/CVF Conference on Computer Vision and Pattern Recognition}}. \bibinfo{pages}{300--309}.
\newblock


\bibitem[Liu et~al\mbox{.}(2024)]%
        {survey_3D_Liu24}
\bibfield{author}{\bibinfo{person}{Jian Liu}, \bibinfo{person}{Xiaoshui Huang}, \bibinfo{person}{Tianyu Huang}, \bibinfo{person}{Lu Chen}, \bibinfo{person}{Yuenan Hou}, \bibinfo{person}{Shixiang Tang}, \bibinfo{person}{Ziwei Liu}, \bibinfo{person}{Wanli Ouyang}, \bibinfo{person}{Wangmeng Zuo}, \bibinfo{person}{Junjun Jiang}, {and} \bibinfo{person}{Xianming Liu}.} \bibinfo{year}{2024}\natexlab{}.
\newblock \showarticletitle{A Comprehensive Survey on 3D Content Generation}.
\newblock \bibinfo{journal}{\emph{CoRR}}  \bibinfo{volume}{abs/2402.01166} (\bibinfo{year}{2024}).
\newblock


\bibitem[Liu et~al\mbox{.}(2023c)]%
        {stylenerf}
\bibfield{author}{\bibinfo{person}{Kunhao Liu}, \bibinfo{person}{Fangneng Zhan}, \bibinfo{person}{Yiwen Chen}, \bibinfo{person}{Jiahui Zhang}, \bibinfo{person}{Yingchen Yu}, \bibinfo{person}{Abdulmotaleb El~Saddik}, \bibinfo{person}{Shijian Lu}, {and} \bibinfo{person}{Eric~P Xing}.} \bibinfo{year}{2023}\natexlab{c}.
\newblock \showarticletitle{{StyleRF}: Zero-shot {3D} Style Transfer of Neural Radiance Fields}. In \bibinfo{booktitle}{\emph{Proceedings of the IEEE/CVF Conference on Computer Vision and Pattern Recognition}}. \bibinfo{pages}{8338--8348}.
\newblock


\bibitem[Liu et~al\mbox{.}(2023b)]%
        {zero123}
\bibfield{author}{\bibinfo{person}{Ruoshi Liu}, \bibinfo{person}{Rundi Wu}, \bibinfo{person}{Basile Van~Hoorick}, \bibinfo{person}{Pavel Tokmakov}, \bibinfo{person}{Sergey Zakharov}, {and} \bibinfo{person}{Carl Vondrick}.} \bibinfo{year}{2023}\natexlab{b}.
\newblock \showarticletitle{Zero-1-to-3: Zero-shot one image to {3D} object}. In \bibinfo{booktitle}{\emph{Proceedings of the IEEE/CVF International Conference on Computer Vision}}. \bibinfo{pages}{9298--9309}.
\newblock


\bibitem[Liu et~al\mbox{.}(2021)]%
        {Edit_NeRF}
\bibfield{author}{\bibinfo{person}{Steven Liu}, \bibinfo{person}{Xiuming Zhang}, \bibinfo{person}{Zhoutong Zhang}, \bibinfo{person}{Richard Zhang}, \bibinfo{person}{Jun-Yan Zhu}, {and} \bibinfo{person}{Bryan Russell}.} \bibinfo{year}{2021}\natexlab{}.
\newblock \showarticletitle{Editing conditional radiance fields}. In \bibinfo{booktitle}{\emph{Proceedings of the IEEE/CVF International Conference on Computer Vision}}. \bibinfo{pages}{5773--5783}.
\newblock


\bibitem[Liu et~al\mbox{.}(2023a)]%
        {SyncDreamer}
\bibfield{author}{\bibinfo{person}{Yuan Liu}, \bibinfo{person}{Cheng Lin}, \bibinfo{person}{Zijiao Zeng}, \bibinfo{person}{Xiaoxiao Long}, \bibinfo{person}{Lingjie Liu}, \bibinfo{person}{Taku Komura}, {and} \bibinfo{person}{Wenping Wang}.} \bibinfo{year}{2023}\natexlab{a}.
\newblock \showarticletitle{{SyncDreamer}: Generating Multiview-consistent Images from a Single-view Image}.
\newblock \bibinfo{journal}{\emph{CoRR}}  \bibinfo{volume}{abs/2309.03453} (\bibinfo{year}{2023}).
\newblock
\urldef\tempurl%
\url{https://doi.org/10.48550/arXiv.2309.03453}
\showURL{%
\tempurl}


\bibitem[Lun et~al\mbox{.}(2017)]%
        {lun2017_3d_pc}
\bibfield{author}{\bibinfo{person}{Zhaoliang Lun}, \bibinfo{person}{Matheus Gadelha}, \bibinfo{person}{Evangelos Kalogerakis}, \bibinfo{person}{Subhransu Maji}, {and} \bibinfo{person}{Rui Wang}.} \bibinfo{year}{2017}\natexlab{}.
\newblock \showarticletitle{{3D} shape reconstruction from sketches via multi-view convolutional networks}. In \bibinfo{booktitle}{\emph{International Conference on 3D Vision (3DV)}}. \bibinfo{pages}{67--77}.
\newblock


\bibitem[Melas-Kyriazi et~al\mbox{.}(2023)]%
        {Realfusion}
\bibfield{author}{\bibinfo{person}{Luke Melas-Kyriazi}, \bibinfo{person}{Iro Laina}, \bibinfo{person}{Christian Rupprecht}, {and} \bibinfo{person}{Andrea Vedaldi}.} \bibinfo{year}{2023}\natexlab{}.
\newblock \showarticletitle{{RealFusion}: 360deg reconstruction of any object from a single image}. In \bibinfo{booktitle}{\emph{Proceedings of the IEEE/CVF Conference on Computer Vision and Pattern Recognition}}. \bibinfo{pages}{8446--8455}.
\newblock


\bibitem[Metzer et~al\mbox{.}(2023)]%
        {latent_nerf}
\bibfield{author}{\bibinfo{person}{Gal Metzer}, \bibinfo{person}{Elad Richardson}, \bibinfo{person}{Or Patashnik}, \bibinfo{person}{Raja Giryes}, {and} \bibinfo{person}{Daniel Cohen-Or}.} \bibinfo{year}{2023}\natexlab{}.
\newblock \showarticletitle{Latent-{NeRF} for shape-guided generation of {3D} shapes and textures}. In \bibinfo{booktitle}{\emph{Proceedings of the IEEE/CVF Conference on Computer Vision and Pattern Recognition}}. \bibinfo{pages}{12663--12673}.
\newblock


\bibitem[{Midjournal.}(2022)]%
        {Midjournal}
\bibfield{author}{\bibinfo{person}{{Midjournal.}}} \bibinfo{year}{2022}\natexlab{}.
\newblock \bibinfo{booktitle}{\emph{Midjournal}}.
\newblock
\urldef\tempurl%
\url{https://www.midjourney.com/}
\showURL{%
\tempurl}


\bibitem[Mikaeili et~al\mbox{.}(2023)]%
        {Sked}
\bibfield{author}{\bibinfo{person}{Aryan Mikaeili}, \bibinfo{person}{Or Perel}, \bibinfo{person}{Mehdi Safaee}, \bibinfo{person}{Daniel Cohen-Or}, {and} \bibinfo{person}{Ali Mahdavi-Amiri}.} \bibinfo{year}{2023}\natexlab{}.
\newblock \showarticletitle{{SKED}: Sketch-guided text-based {3D} editing}. In \bibinfo{booktitle}{\emph{Proceedings of the IEEE/CVF International Conference on Computer Vision}}. \bibinfo{pages}{14607--14619}.
\newblock


\bibitem[Mou et~al\mbox{.}(2024)]%
        {T2I-Adapter}
\bibfield{author}{\bibinfo{person}{Chong Mou}, \bibinfo{person}{Xintao Wang}, \bibinfo{person}{Liangbin Xie}, \bibinfo{person}{Yanze Wu}, \bibinfo{person}{Jian Zhang}, \bibinfo{person}{Zhongang Qi}, {and} \bibinfo{person}{Ying Shan}.} \bibinfo{year}{2024}\natexlab{}.
\newblock \showarticletitle{T2I-Adapter: Learning Adapters to Dig Out More Controllable Ability for Text-to-Image Diffusion Models}. In \bibinfo{booktitle}{\emph{Conference on Artificial Intelligence}}. \bibinfo{pages}{4296--4304}.
\newblock


\bibitem[Nguyen{-}Phuoc et~al\mbox{.}(2022)]%
        {SNeRF}
\bibfield{author}{\bibinfo{person}{Thu Nguyen{-}Phuoc}, \bibinfo{person}{Feng Liu}, {and} \bibinfo{person}{Lei Xiao}.} \bibinfo{year}{2022}\natexlab{}.
\newblock \showarticletitle{SNeRF: stylized neural implicit representations for {3D} scenes}.
\newblock \bibinfo{journal}{\emph{ACM Transactions on Graphics}} \bibinfo{volume}{41}, \bibinfo{number}{4} (\bibinfo{year}{2022}), \bibinfo{pages}{142:1--142:11}.
\newblock


\bibitem[Poole et~al\mbox{.}(2023)]%
        {Dreamfusion}
\bibfield{author}{\bibinfo{person}{Ben Poole}, \bibinfo{person}{Ajay Jain}, \bibinfo{person}{Jonathan~T. Barron}, {and} \bibinfo{person}{Ben Mildenhall}.} \bibinfo{year}{2023}\natexlab{}.
\newblock \showarticletitle{{DreamFusion}: Text-to-{3D} using {2D} Diffusion}. In \bibinfo{booktitle}{\emph{International Conference on Learning Representations}}.
\newblock


\bibitem[Qian et~al\mbox{.}(2023)]%
        {magic123}
\bibfield{author}{\bibinfo{person}{Guocheng Qian}, \bibinfo{person}{Jinjie Mai}, \bibinfo{person}{Abdullah Hamdi}, \bibinfo{person}{Jian Ren}, \bibinfo{person}{Aliaksandr Siarohin}, \bibinfo{person}{Bing Li}, \bibinfo{person}{Hsin{-}Ying Lee}, \bibinfo{person}{Ivan Skorokhodov}, \bibinfo{person}{Peter Wonka}, \bibinfo{person}{Sergey Tulyakov}, {and} \bibinfo{person}{Bernard Ghanem}.} \bibinfo{year}{2023}\natexlab{}.
\newblock \showarticletitle{Magic123: One Image to High-Quality {3D} Object Generation Using Both {2D} and {3D} Diffusion Priors}.
\newblock \bibinfo{journal}{\emph{CoRR}}  \bibinfo{volume}{abs/2306.17843} (\bibinfo{year}{2023}).
\newblock


\bibitem[Qiu et~al\mbox{.}(2023)]%
        {richdreamer}
\bibfield{author}{\bibinfo{person}{Lingteng Qiu}, \bibinfo{person}{Guanying Chen}, \bibinfo{person}{Xiaodong Gu}, \bibinfo{person}{Qi Zuo}, \bibinfo{person}{Mutian Xu}, \bibinfo{person}{Yushuang Wu}, \bibinfo{person}{Weihao Yuan}, \bibinfo{person}{Zilong Dong}, \bibinfo{person}{Liefeng Bo}, {and} \bibinfo{person}{Xiaoguang Han}.} \bibinfo{year}{2023}\natexlab{}.
\newblock \showarticletitle{{RichDreamer}: A Generalizable Normal-Depth Diffusion Model for Detail Richness in Text-to-{3D}}.
\newblock \bibinfo{journal}{\emph{arXiv preprint arXiv:2311.16918}} (\bibinfo{year}{2023}).
\newblock


\bibitem[Radford et~al\mbox{.}(2021)]%
        {CLIP}
\bibfield{author}{\bibinfo{person}{Alec Radford}, \bibinfo{person}{Jong~Wook Kim}, \bibinfo{person}{Chris Hallacy}, \bibinfo{person}{Aditya Ramesh}, \bibinfo{person}{Gabriel Goh}, \bibinfo{person}{Sandhini Agarwal}, \bibinfo{person}{Girish Sastry}, \bibinfo{person}{Amanda Askell}, \bibinfo{person}{Pamela Mishkin}, \bibinfo{person}{Jack Clark}, {et~al\mbox{.}}} \bibinfo{year}{2021}\natexlab{}.
\newblock \showarticletitle{Learning transferable visual models from natural language supervision}. In \bibinfo{booktitle}{\emph{International conference on machine learning}}. \bibinfo{pages}{8748--8763}.
\newblock


\bibitem[Rombach et~al\mbox{.}(2022)]%
        {stable_diffusion}
\bibfield{author}{\bibinfo{person}{Robin Rombach}, \bibinfo{person}{Andreas Blattmann}, \bibinfo{person}{Dominik Lorenz}, \bibinfo{person}{Patrick Esser}, {and} \bibinfo{person}{Bj{\"o}rn Ommer}.} \bibinfo{year}{2022}\natexlab{}.
\newblock \showarticletitle{High-resolution image synthesis with latent diffusion models}. In \bibinfo{booktitle}{\emph{Proceedings of the IEEE/CVF Conference on Computer Vision and Pattern Recognition}}. \bibinfo{pages}{10684--10695}.
\newblock


\bibitem[Ruiz et~al\mbox{.}(2023)]%
        {Dreambooth}
\bibfield{author}{\bibinfo{person}{Nataniel Ruiz}, \bibinfo{person}{Yuanzhen Li}, \bibinfo{person}{Varun Jampani}, \bibinfo{person}{Yael Pritch}, \bibinfo{person}{Michael Rubinstein}, {and} \bibinfo{person}{Kfir Aberman}.} \bibinfo{year}{2023}\natexlab{}.
\newblock \showarticletitle{{DreamBooth}: Fine tuning text-to-image diffusion models for subject-driven generation}. In \bibinfo{booktitle}{\emph{Proceedings of the IEEE/CVF Conference on Computer Vision and Pattern Recognition}}. \bibinfo{pages}{22500--22510}.
\newblock


\bibitem[Sella et~al\mbox{.}(2023)]%
        {vox-e}
\bibfield{author}{\bibinfo{person}{Etai Sella}, \bibinfo{person}{Gal Fiebelman}, \bibinfo{person}{Peter Hedman}, {and} \bibinfo{person}{Hadar Averbuch-Elor}.} \bibinfo{year}{2023}\natexlab{}.
\newblock \showarticletitle{Vox-{E}: Text-guided voxel editing of {3D} objects}. In \bibinfo{booktitle}{\emph{Proceedings of the IEEE/CVF International Conference on Computer Vision}}. \bibinfo{pages}{430--440}.
\newblock


\bibitem[Shi et~al\mbox{.}(2023a)]%
        {zero123++}
\bibfield{author}{\bibinfo{person}{Ruoxi Shi}, \bibinfo{person}{Hansheng Chen}, \bibinfo{person}{Zhuoyang Zhang}, \bibinfo{person}{Minghua Liu}, \bibinfo{person}{Chao Xu}, \bibinfo{person}{Xinyue Wei}, \bibinfo{person}{Linghao Chen}, \bibinfo{person}{Chong Zeng}, {and} \bibinfo{person}{Hao Su}.} \bibinfo{year}{2023}\natexlab{a}.
\newblock \showarticletitle{Zero123++: a Single Image to Consistent Multi-view Diffusion Base Model}.
\newblock \bibinfo{journal}{\emph{CoRR}}  \bibinfo{volume}{abs/2310.15110} (\bibinfo{year}{2023}).
\newblock


\bibitem[Shi et~al\mbox{.}(2023b)]%
        {mvdream}
\bibfield{author}{\bibinfo{person}{Yichun Shi}, \bibinfo{person}{Peng Wang}, \bibinfo{person}{Jianglong Ye}, \bibinfo{person}{Mai Long}, \bibinfo{person}{Kejie Li}, {and} \bibinfo{person}{Xiao Yang}.} \bibinfo{year}{2023}\natexlab{b}.
\newblock \showarticletitle{{MVDream}: Multi-view diffusion for {3D} generation}.
\newblock \bibinfo{journal}{\emph{arXiv preprint arXiv:2308.16512}} (\bibinfo{year}{2023}).
\newblock


\bibitem[Somraj(2020)]%
        {posewarping}
\bibfield{author}{\bibinfo{person}{Nagabhushan Somraj}.} \bibinfo{year}{2020}\natexlab{}.
\newblock \bibinfo{title}{Pose-Warping for View Synthesis / {DIBR}}.
\newblock
\newblock
\urldef\tempurl%
\url{https://github.com/NagabhushanSN95/Pose-Warping}
\showURL{%
\tempurl}


\bibitem[Sun et~al\mbox{.}(2024)]%
        {survey_3D_Sun0024}
\bibfield{author}{\bibinfo{person}{Jia{-}Mu Sun}, \bibinfo{person}{Tong Wu}, {and} \bibinfo{person}{Lin Gao}.} \bibinfo{year}{2024}\natexlab{}.
\newblock \showarticletitle{Recent advances in implicit representation-based 3D shape generation}.
\newblock \bibinfo{journal}{\emph{Vis. Intell.}} \bibinfo{volume}{2}, \bibinfo{number}{1} (\bibinfo{year}{2024}).
\newblock


\bibitem[Sun et~al\mbox{.}(2023)]%
        {Dreamcraft3d}
\bibfield{author}{\bibinfo{person}{Jingxiang Sun}, \bibinfo{person}{Bo Zhang}, \bibinfo{person}{Ruizhi Shao}, \bibinfo{person}{Lizhen Wang}, \bibinfo{person}{Wen Liu}, \bibinfo{person}{Zhenda Xie}, {and} \bibinfo{person}{Yebin Liu}.} \bibinfo{year}{2023}\natexlab{}.
\newblock \showarticletitle{{Dreamcraft3D}: Hierarchical {3D} generation with bootstrapped diffusion prior}.
\newblock \bibinfo{journal}{\emph{arXiv preprint arXiv:2310.16818}} (\bibinfo{year}{2023}).
\newblock


\bibitem[Tang et~al\mbox{.}(2023a)]%
        {Dreamgaussian}
\bibfield{author}{\bibinfo{person}{Jiaxiang Tang}, \bibinfo{person}{Jiawei Ren}, \bibinfo{person}{Hang Zhou}, \bibinfo{person}{Ziwei Liu}, {and} \bibinfo{person}{Gang Zeng}.} \bibinfo{year}{2023}\natexlab{a}.
\newblock \showarticletitle{{DreamGaussian}: Generative {Gaussian} splatting for efficient {3D} content creation}.
\newblock \bibinfo{journal}{\emph{arXiv preprint arXiv:2309.16653}} (\bibinfo{year}{2023}).
\newblock


\bibitem[Tang et~al\mbox{.}(2023b)]%
        {Make-It-3D}
\bibfield{author}{\bibinfo{person}{Junshu Tang}, \bibinfo{person}{Tengfei Wang}, \bibinfo{person}{Bo Zhang}, \bibinfo{person}{Ting Zhang}, \bibinfo{person}{Ran Yi}, \bibinfo{person}{Lizhuang Ma}, {and} \bibinfo{person}{Dong Chen}.} \bibinfo{year}{2023}\natexlab{b}.
\newblock \showarticletitle{{Make-It-3D}: High-fidelity {3D} Creation from A Single Image with Diffusion Prior}. In \bibinfo{booktitle}{\emph{Proceedings of the IEEE/CVF International Conference on Computer Vision}}. \bibinfo{pages}{22819--22829}.
\newblock


\bibitem[Wang et~al\mbox{.}(2022a)]%
        {clip-nerf}
\bibfield{author}{\bibinfo{person}{Can Wang}, \bibinfo{person}{Menglei Chai}, \bibinfo{person}{Mingming He}, \bibinfo{person}{Dongdong Chen}, {and} \bibinfo{person}{Jing Liao}.} \bibinfo{year}{2022}\natexlab{a}.
\newblock \showarticletitle{{CLIP-NeRF}: Text-and-image driven manipulation of neural radiance fields}. In \bibinfo{booktitle}{\emph{Proceedings of the IEEE/CVF Conference on Computer Vision and Pattern Recognition}}. \bibinfo{pages}{3835--3844}.
\newblock


\bibitem[Wang et~al\mbox{.}(2023b)]%
        {nerf-art}
\bibfield{author}{\bibinfo{person}{Can Wang}, \bibinfo{person}{Ruixiang Jiang}, \bibinfo{person}{Menglei Chai}, \bibinfo{person}{Mingming He}, \bibinfo{person}{Dongdong Chen}, {and} \bibinfo{person}{Jing Liao}.} \bibinfo{year}{2023}\natexlab{b}.
\newblock \showarticletitle{{NeRF-Art}: Text-driven neural radiance fields stylization}.
\newblock \bibinfo{journal}{\emph{IEEE Transactions on Visualization and Computer Graphics}} (\bibinfo{year}{2023}).
\newblock


\bibitem[Wang et~al\mbox{.}(2023a)]%
        {SJC}
\bibfield{author}{\bibinfo{person}{Haochen Wang}, \bibinfo{person}{Xiaodan Du}, \bibinfo{person}{Jiahao Li}, \bibinfo{person}{Raymond~A Yeh}, {and} \bibinfo{person}{Greg Shakhnarovich}.} \bibinfo{year}{2023}\natexlab{a}.
\newblock \showarticletitle{Score jacobian chaining: Lifting pretrained {2D} diffusion models for {3D} generation}. In \bibinfo{booktitle}{\emph{Proceedings of the IEEE/CVF Conference on Computer Vision and Pattern Recognition}}. \bibinfo{pages}{12619--12629}.
\newblock


\bibitem[Wang et~al\mbox{.}(2022b)]%
        {wang2022_3d_pc}
\bibfield{author}{\bibinfo{person}{Jiayun Wang}, \bibinfo{person}{Jierui Lin}, \bibinfo{person}{Qian Yu}, \bibinfo{person}{Runtao Liu}, \bibinfo{person}{Yubei Chen}, {and} \bibinfo{person}{Stella~X Yu}.} \bibinfo{year}{2022}\natexlab{b}.
\newblock \showarticletitle{{3D} shape reconstruction from free-hand sketches}. In \bibinfo{booktitle}{\emph{European Conference on Computer Vision}}. \bibinfo{pages}{184--202}.
\newblock


\bibitem[Wang et~al\mbox{.}(2018)]%
        {2018_unsupervised_volumn}
\bibfield{author}{\bibinfo{person}{Lingjing Wang}, \bibinfo{person}{Cheng Qian}, \bibinfo{person}{Jifei Wang}, {and} \bibinfo{person}{Yi Fang}.} \bibinfo{year}{2018}\natexlab{}.
\newblock \showarticletitle{Unsupervised learning of {3D} model reconstruction from hand-drawn sketches}. In \bibinfo{booktitle}{\emph{Proceedings of the 26th ACM international conference on Multimedia}}. \bibinfo{pages}{1820--1828}.
\newblock


\bibitem[Wang and Shi(2023)]%
        {Imagedream}
\bibfield{author}{\bibinfo{person}{Peng Wang} {and} \bibinfo{person}{Yichun Shi}.} \bibinfo{year}{2023}\natexlab{}.
\newblock \showarticletitle{{ImageDream}: Image-Prompt Multi-view Diffusion for {3D} Generation}.
\newblock \bibinfo{journal}{\emph{arXiv preprint arXiv:2312.02201}} (\bibinfo{year}{2023}).
\newblock


\bibitem[Wang et~al\mbox{.}(2023d)]%
        {Rodin}
\bibfield{author}{\bibinfo{person}{Tengfei Wang}, \bibinfo{person}{Bo Zhang}, \bibinfo{person}{Ting Zhang}, \bibinfo{person}{Shuyang Gu}, \bibinfo{person}{Jianmin Bao}, \bibinfo{person}{Tadas Baltrusaitis}, \bibinfo{person}{Jingjing Shen}, \bibinfo{person}{Dong Chen}, \bibinfo{person}{Fang Wen}, \bibinfo{person}{Qifeng Chen}, {et~al\mbox{.}}} \bibinfo{year}{2023}\natexlab{d}.
\newblock \showarticletitle{{RODIN}: A generative model for sculpting {3D} digital avatars using diffusion}. In \bibinfo{booktitle}{\emph{Proceedings of the IEEE/CVF Conference on Computer Vision and Pattern Recognition}}. \bibinfo{pages}{4563--4573}.
\newblock


\bibitem[Wang et~al\mbox{.}(2023c)]%
        {ProlificDreamer}
\bibfield{author}{\bibinfo{person}{Zhengyi Wang}, \bibinfo{person}{Cheng Lu}, \bibinfo{person}{Yikai Wang}, \bibinfo{person}{Fan Bao}, \bibinfo{person}{Chongxuan Li}, \bibinfo{person}{Hang Su}, {and} \bibinfo{person}{Jun Zhu}.} \bibinfo{year}{2023}\natexlab{c}.
\newblock \showarticletitle{ProlificDreamer: High-Fidelity and Diverse Text-to-3D Generation with Variational Score Distillation}. In \bibinfo{booktitle}{\emph{Advances in Neural Information Processing Systems}}.
\newblock


\bibitem[Wu et~al\mbox{.}(2023)]%
        {Hyperdreamer}
\bibfield{author}{\bibinfo{person}{Tong Wu}, \bibinfo{person}{Zhibing Li}, \bibinfo{person}{Shuai Yang}, \bibinfo{person}{Pan Zhang}, \bibinfo{person}{Xingang Pan}, \bibinfo{person}{Jiaqi Wang}, \bibinfo{person}{Dahua Lin}, {and} \bibinfo{person}{Ziwei Liu}.} \bibinfo{year}{2023}\natexlab{}.
\newblock \showarticletitle{{HyperDreamer}: Hyper-Realistic {3D} Content Generation and Editing from a Single Image}. In \bibinfo{booktitle}{\emph{SIGGRAPH Asia 2023 Conference Papers}}. \bibinfo{pages}{1--10}.
\newblock


\bibitem[Xia and Xue(2024)]%
        {survey_3D_XiaX24}
\bibfield{author}{\bibinfo{person}{Weihao Xia} {and} \bibinfo{person}{Jing{-}Hao Xue}.} \bibinfo{year}{2024}\natexlab{}.
\newblock \showarticletitle{A Survey on Deep Generative 3D-aware Image Synthesis}.
\newblock \bibinfo{journal}{\emph{{ACM} Comput. Surv.}} \bibinfo{volume}{56}, \bibinfo{number}{4} (\bibinfo{year}{2024}), \bibinfo{pages}{90:1--90:34}.
\newblock


\bibitem[Xiang et~al\mbox{.}(2020)]%
        {xiang2020sketch_mesh}
\bibfield{author}{\bibinfo{person}{Nan Xiang}, \bibinfo{person}{Ruibin Wang}, \bibinfo{person}{Tao Jiang}, \bibinfo{person}{Li Wang}, \bibinfo{person}{Yanran Li}, \bibinfo{person}{Xiaosong Yang}, {and} \bibinfo{person}{Jianjun Zhang}.} \bibinfo{year}{2020}\natexlab{}.
\newblock \showarticletitle{Sketch-based modeling with a differentiable renderer}.
\newblock \bibinfo{journal}{\emph{Computer Animation and Virtual Worlds}} \bibinfo{volume}{31}, \bibinfo{number}{4-5} (\bibinfo{year}{2020}), \bibinfo{pages}{e1939}.
\newblock


\bibitem[Xu and Harada(2022)]%
        {xu2022deforming}
\bibfield{author}{\bibinfo{person}{Tianhan Xu} {and} \bibinfo{person}{Tatsuya Harada}.} \bibinfo{year}{2022}\natexlab{}.
\newblock \showarticletitle{Deforming radiance fields with cages}. In \bibinfo{booktitle}{\emph{European Conference on Computer Vision}}. \bibinfo{pages}{159--175}.
\newblock


\bibitem[Yang et~al\mbox{.}(2022)]%
        {Neumesh}
\bibfield{author}{\bibinfo{person}{Bangbang Yang}, \bibinfo{person}{Chong Bao}, \bibinfo{person}{Junyi Zeng}, \bibinfo{person}{Hujun Bao}, \bibinfo{person}{Yinda Zhang}, \bibinfo{person}{Zhaopeng Cui}, {and} \bibinfo{person}{Guofeng Zhang}.} \bibinfo{year}{2022}\natexlab{}.
\newblock \showarticletitle{{NeuMesh}: Learning disentangled neural mesh-based implicit field for geometry and texture editing}. In \bibinfo{booktitle}{\emph{European Conference on Computer Vision}}. \bibinfo{pages}{597--614}.
\newblock


\bibitem[Yang et~al\mbox{.}(2021)]%
        {yang2021learning}
\bibfield{author}{\bibinfo{person}{Bangbang Yang}, \bibinfo{person}{Yinda Zhang}, \bibinfo{person}{Yinghao Xu}, \bibinfo{person}{Yijin Li}, \bibinfo{person}{Han Zhou}, \bibinfo{person}{Hujun Bao}, \bibinfo{person}{Guofeng Zhang}, {and} \bibinfo{person}{Zhaopeng Cui}.} \bibinfo{year}{2021}\natexlab{}.
\newblock \showarticletitle{Learning object-compositional neural radiance field for editable scene rendering}. In \bibinfo{booktitle}{\emph{Proceedings of the IEEE/CVF International Conference on Computer Vision}}. \bibinfo{pages}{13779--13788}.
\newblock


\bibitem[Yuan et~al\mbox{.}(2022)]%
        {nerf_editing}
\bibfield{author}{\bibinfo{person}{Yu-Jie Yuan}, \bibinfo{person}{Yang-Tian Sun}, \bibinfo{person}{Yu-Kun Lai}, \bibinfo{person}{Yuewen Ma}, \bibinfo{person}{Rongfei Jia}, {and} \bibinfo{person}{Lin Gao}.} \bibinfo{year}{2022}\natexlab{}.
\newblock \showarticletitle{Nerf-editing: geometry editing of neural radiance fields}. In \bibinfo{booktitle}{\emph{Proceedings of the IEEE/CVF Conference on Computer Vision and Pattern Recognition}}. \bibinfo{pages}{18353--18364}.
\newblock


\bibitem[Zeleznik et~al\mbox{.}(2006)]%
        {zeleznik2006sketch}
\bibfield{author}{\bibinfo{person}{Robert~C Zeleznik}, \bibinfo{person}{Kenneth~P Herndon}, {and} \bibinfo{person}{John~F Hughes}.} \bibinfo{year}{2006}\natexlab{}.
\newblock \showarticletitle{{SKETCH}: An interface for sketching {3D} scenes}.
\newblock In \bibinfo{booktitle}{\emph{ACM SIGGRAPH 2006 Courses}}. \bibinfo{pages}{9--es}.
\newblock


\bibitem[Zeng et~al\mbox{.}(2023)]%
        {IPDreamer}
\bibfield{author}{\bibinfo{person}{Bohan Zeng}, \bibinfo{person}{Shanglin Li}, \bibinfo{person}{Yutang Feng}, \bibinfo{person}{Hong Li}, \bibinfo{person}{Sicheng Gao}, \bibinfo{person}{Jiaming Liu}, \bibinfo{person}{Huaxia Li}, \bibinfo{person}{Xu Tang}, \bibinfo{person}{Jianzhuang Liu}, {and} \bibinfo{person}{Baochang Zhang}.} \bibinfo{year}{2023}\natexlab{}.
\newblock \showarticletitle{{IPDreamer}: Appearance-Controllable {3D} Object Generation with Image Prompts}.
\newblock \bibinfo{journal}{\emph{arXiv preprint arXiv:2310.05375}} (\bibinfo{year}{2023}).
\newblock


\bibitem[Zhang et~al\mbox{.}(2021b)]%
        {editable_nerf}
\bibfield{author}{\bibinfo{person}{Jiakai Zhang}, \bibinfo{person}{Xinhang Liu}, \bibinfo{person}{Xinyi Ye}, \bibinfo{person}{Fuqiang Zhao}, \bibinfo{person}{Yanshun Zhang}, \bibinfo{person}{Minye Wu}, \bibinfo{person}{Yingliang Zhang}, \bibinfo{person}{Lan Xu}, {and} \bibinfo{person}{Jingyi Yu}.} \bibinfo{year}{2021}\natexlab{b}.
\newblock \showarticletitle{Editable free-viewpoint video using a layered neural representation}.
\newblock \bibinfo{journal}{\emph{ACM Transactions on Graphics}} \bibinfo{volume}{40}, \bibinfo{number}{4} (\bibinfo{year}{2021}), \bibinfo{pages}{149:1--149:18}.
\newblock


\bibitem[Zhang et~al\mbox{.}(2023)]%
        {controlnet}
\bibfield{author}{\bibinfo{person}{Lvmin Zhang}, \bibinfo{person}{Anyi Rao}, {and} \bibinfo{person}{Maneesh Agrawala}.} \bibinfo{year}{2023}\natexlab{}.
\newblock \showarticletitle{Adding conditional control to text-to-image diffusion models}. In \bibinfo{booktitle}{\emph{Proceedings of the IEEE/CVF International Conference on Computer Vision}}. \bibinfo{pages}{3836--3847}.
\newblock


\bibitem[Zhang et~al\mbox{.}(2021a)]%
        {zhang2021sketch2model}
\bibfield{author}{\bibinfo{person}{Song-Hai Zhang}, \bibinfo{person}{Yuan-Chen Guo}, {and} \bibinfo{person}{Qing-Wen Gu}.} \bibinfo{year}{2021}\natexlab{a}.
\newblock \showarticletitle{Sketch2model: View-aware {3D} modeling from single free-hand sketches}. In \bibinfo{booktitle}{\emph{Proceedings of the IEEE/CVF Conference on Computer Vision and Pattern Recognition}}. \bibinfo{pages}{6012--6021}.
\newblock


\bibitem[Zheng et~al\mbox{.}(2023)]%
        {zheng_diffusion_SDF}
\bibfield{author}{\bibinfo{person}{Xin{-}Yang Zheng}, \bibinfo{person}{Hao Pan}, \bibinfo{person}{Peng{-}Shuai Wang}, \bibinfo{person}{Xin Tong}, \bibinfo{person}{Yang Liu}, {and} \bibinfo{person}{Heung{-}Yeung Shum}.} \bibinfo{year}{2023}\natexlab{}.
\newblock \showarticletitle{Locally Attentional {SDF} Diffusion for Controllable {3D} Shape Generation}.
\newblock \bibinfo{journal}{\emph{ACM Transactions on Graphics}} \bibinfo{volume}{42}, \bibinfo{number}{4} (\bibinfo{year}{2023}), \bibinfo{pages}{91:1--91:13}.
\newblock


\bibitem[Zhong et~al\mbox{.}(2020a)]%
        {zhong2020deep_data_aug}
\bibfield{author}{\bibinfo{person}{Yue Zhong}, \bibinfo{person}{Yulia Gryaditskaya}, \bibinfo{person}{Honggang Zhang}, {and} \bibinfo{person}{Yi-Zhe Song}.} \bibinfo{year}{2020}\natexlab{a}.
\newblock \showarticletitle{Deep sketch-based modeling: Tips and tricks}. In \bibinfo{booktitle}{\emph{2020 International Conference on 3D Vision (3DV)}}. \bibinfo{pages}{543--552}.
\newblock


\bibitem[Zhong et~al\mbox{.}(2020b)]%
        {zhong2020towards}
\bibfield{author}{\bibinfo{person}{Yue Zhong}, \bibinfo{person}{Yonggang Qi}, \bibinfo{person}{Yulia Gryaditskaya}, \bibinfo{person}{Honggang Zhang}, {and} \bibinfo{person}{Yi-Zhe Song}.} \bibinfo{year}{2020}\natexlab{b}.
\newblock \showarticletitle{Towards practical sketch-based {3D} shape generation: The role of professional sketches}.
\newblock \bibinfo{journal}{\emph{IEEE Transactions on Circuits and Systems for Video Technology}} \bibinfo{volume}{31}, \bibinfo{number}{9} (\bibinfo{year}{2020}), \bibinfo{pages}{3518--3528}.
\newblock


\bibitem[Zhuang et~al\mbox{.}(2023)]%
        {Dreameditor}
\bibfield{author}{\bibinfo{person}{Jingyu Zhuang}, \bibinfo{person}{Chen Wang}, \bibinfo{person}{Liang Lin}, \bibinfo{person}{Lingjie Liu}, {and} \bibinfo{person}{Guanbin Li}.} \bibinfo{year}{2023}\natexlab{}.
\newblock \showarticletitle{{DreamEditor}: Text-driven {3D} scene editing with neural fields}. In \bibinfo{booktitle}{\emph{SIGGRAPH Asia 2023 Conference Papers}}. \bibinfo{pages}{1--10}.
\newblock


\end{thebibliography}
